\documentclass[fleqn,usenatbib]{mnras}
\usepackage{newtxtext,newtxmath}
\usepackage{longtable}

\usepackage[T1]{fontenc}

\DeclareRobustCommand{\VAN}[3]{#2}
\let\VANthebibliography\thebibliography
\def\thebibliography{\DeclareRobustCommand{\VAN}[3]{##3}\VANthebibliography}

\usepackage{graphicx}
\usepackage{subcaption} 
\usepackage[version=4,arrows=pgf-filled,
textfontname=sffamily,
mathfontname=mathsf]{mhchem}

\usepackage{xcolor}
\usepackage{pifont}
\usepackage{gensymb}
\usepackage{physics}   
\usepackage{wasysym}

\usepackage{txfonts}
\title{Hot and cloudy: High temperature clouds in super-Earths and sub-Neptunes}

   \author[L.J. Janssen et al.]{
L. J. Janssen,$^{1}$\thanks{E-mail: ljanssen@strw.leidenuniv.nl},
          Y. Miguel$^{1,2}$,
          M. Min$^{2}$,
          H. Huang$^{3}$,
          M. Zilinskas$^{2,4}$,
          C.P.A. van Buchem$^{1}$
\\
$^{1}$Leiden Observatory, Einsteinweg 55, 2333 CC Leiden\\
$^{2}$SRON Netherlands Institute for Space Research, Niels Bohrweg 4, 2333 CA Leiden, The Netherlands\\
$^{3}$Department of Astronomy,Tsinghua University, Haidian DS 100084, Beijing, China\\
$^{4}$Jet Propulsion Laboratory, California Institute of Technology, Pasadena, CA, USA}

\date{Accepted XXX. Received YYY; in original form ZZZ}

\pubyear{\the\year{2026}}

\begin{document}
\label{firstpage}
\pagerange{\pageref{firstpage}--\pageref{lastpage}}
\maketitle

\begin{abstract}

JWST observations provide for the first time evidence for an atmosphere on a rocky exoplanet - 55 Cnc e. The atmosphere of 55 Cnc e is hot with $\text{T}_{\text{eq}}>2000$\,K and shows strong variability, for which cloud formation above a molten crust could be one possible explanation. The composition of the atmosphere of 55 Cnc e is still unknown but suggests the presence of volatiles. We have run cloud formation models on a grid of \ce{N}-dominated, \ce{O}-dominated, \ce{C}-dominated and \ce{H}-dominated atmospheres to investigate which type of cloud we could expect on hot super-Earths and hot sub-Neptunes ($1000$\,K $<$ T $<$ $3000$\,K). Our models combine radiative transfer with equilibrium chemistry of the gaseous and condensed phases, vertical mixing of condensable species, sedimentation, nucleation and coagulation. We find that the condensability of species is highly dependent on the oxygen abundance of an atmosphere. Oxygen poor atmospheres can be heated by UV and optical absorbers \ce{PS}, \ce{TiO} and \ce{CN} which create temperature inversions. These inhibit condensation. Oxygen rich atmospheres are colder without temperature inversions, and are therefore more favourable environments for cloud formation. The major expected cloud component in \ce{O}-dominated atmospheres with solar refractory abundance is \ce{TiO2(s)}. Spectral features of clouds in these worlds are stronger in transmission than in emission, in particular at short wavelengths. We find a lack of optical data of solid species in comparison to the variety of stable cloud components which can form on hot, rocky planets. 
\end{abstract}




\section{Introduction}
Short period planets - $P<100$ days, which are smaller than Neptune are numerous \citep{Fulton2017,Bean2021}. A lack of planets around $1.8$\,R$_{\oplus}$ has been observed dividing these small objects into two categories: Smaller planets $<1.8$\,R$_{\oplus}$ (super-Earths) and planets of $2-3$\,R$_{\oplus}$ (sub-Neptunes) \citep[e.g.][]{Fulton2017,Izidoro2022,Burn2024}. The reason for the lack of planets between $1.5-2$\,R$_{\oplus}$ - the so called Radius Valley - is debated. Both, atmospheric mass loss driven by internal heating of the planet and/or by XUV flux from the star, as well as a composition gap between the two populations have been proposed as a potential cause \citet{Owen&Wu2013,Fulton2017,Ginzburg2018,Luque2022}. These small, hot planets are ideal candidates for studying rocky planet characteristics. Due to the proximity to their host stars, they are bright in thermal emission and have a higher transit probability compared to their temperate counterparts. Hot, rocky planets with surfaces $> 1500$\,K are expected to be molten \citep{vBuchem2023}. We can probe the interiors of these molten planets by probing their atmospheres because material is outgassed from the magma into the atmosphere. Studying outgassed atmospheres of hot, rocky exoplanets can teach us about different evolutionary stages of rocky planets in a more general context. Temperate rocky planets are also expected to go through a molten stage in their evolution before solidifying \citep{Lichtenberg&Miguel2025}.

Observations of small planets are looking more and more promising. \citet{Hu2024} have for the first time shown the existence of an atmosphere on a hot, rocky planet: 55Cnc e. \citet{Zieba2023} rule out a thick \ce{CO2} atmosphere on TRAPPIST-1c and \citet{Ducrot2025} show that a hazy, \ce{CO2} atmosphere is a possibilty for TRAPPIST-1b. \citet{Dang2024} have obtained phase curve observations of the hot super Earth K2-141b. Ariel will allow full phase curve measurements of sub-Neptune sized planets opening up new possibilities for the understanding of their atmospheres in 3D. Additionally, JWST programs such as the Hot Rock Survey are ongoing \citep{August2025,MeierValdes2025}. This program targets nine hot, rocky exoplanets around M-dwarfs to reveal whether such planets can have atmospheres. The JWST Cycle 4 programme includes the Hot Rocky planet survey by \citet{Dang2025}, which will provide us with an overview on which hot, rocky planets can sustain an atmosphere and inform us whether their interiors are fully molten. In the future, the Rocky Worlds DDT JWST program \citep{Redfield2024} promises further observations of hot, rocky planets in emission. 

In such hot, environments, we can still expect some solid and liquid species to be stable. \citet{Wakeford2017} have shown that some high temperature condensates form at temperatures higher than $2000$\,K. In hot, rocky planets the availability of condensable material could be enhanced compared to gas giants because not only volatiles but also refractories such as silicon, iron, potassium, sodium, aluminium and calcium are outgassed in the form of a secondary atmosphere \citep{vBuchem2023,vBuchem2024,Gaillard2014}. If some of these condensable materials form clouds, they could play an important role in controlling the climate on the planet due to clouds' effects on the gas phase composition, on the temperature profile, and on the atmospheric dynamics in planetary atmospheres \citep{Helling&Woitke2006,Powell2024,Carone2023, Bell2024}. Clouds set the planetary albedo, and can in some cases create a green house effect and trap IR radiation on the planet \citep[e.g.][]{2013Heng&Demory,Helling2019,Essack2020}. They affect the atmospheric composition by being sinks for certain gas phase species \citep[e.g.]{Huang2024}. Concerning the observational aspect, clouds can diminish the magnitude of spectral features or generate new features \citep[e.g.][]{Kreidberg2014,Grant2023}.

In this work, we explore which solid and liquid species can be stable in atmospheres of hot, sub-Neptune sized planets, including both super-Earths and sub-Neptunes. We investigate how clouds composed of the identified species affect observations at the relevant wavelengths ($0.1-30$\,$\mu$m) to prepare for the interpretation of upcoming data of current and future missions observing these planets such as JWST, Ariel and PLATO \citep{Gardner2006,Tinetti2018,Rauer2025}.\\

55 Cnc e is a particularly interesting case to explore, since it falls in between what is called a sub-Neptune and a super-Earth \citep{Hu2024}. Strong variability has been detected in multiple observing programs \citep{Dragomir2014,Demory2016,Meier-Valdes2023}, but the cause it still unknown and the possible explanations range from volcanism, to a dust torus to a transient atmosphere \citep{Demory2016,Meier-Valdes2023,Heng2023}.
A recent study by \citet{Loftus2024} suggests that a cyclical pattern between outgassing, cloud formation, and resulting heating of the atmosphere could be another explanation for the strong variability observed on 55 Cnc e. Cloud formation from microphysics on 55 Cnc e had been considered before by \citet{Mahapatra2017}. They compute equilibrium chemistry of the gas phase for a hydrogen-dominated atmosphere and make use of the 1D microphysical cloud model by \citet{Helling&Woitke2006,Helling2008}. They use a temperature structure from retrievals by \citet{Demory2016} and find that the most stable clouds on 55 Cnc e are magnesium- and silicon- oxides. However, atmospheres of rocky planets are not necessarily \ce{H}-dominated. On the contrary, \citet{Heng2025} suggests that the mean molecular weight of the atmosphere increases for smaller planets. Similar conclusions are also drawn by \citet{Cherubim2025}, who show that an \ce{O2}-dominated  atmosphere is the most likely composition for hot, rocky planets. 
If we base the conclusions of potential atmospheres of hot rocky planets on observations, the possibilities for their compositions are numerous. \citet{Zilinskas2025} show that good fits for spectra of the hot super-Earth 55 Cnc e taken by MIRI and NIRSpec JWST, can be nitrogen, phosphorus, carbon, oxygen and/or hydrogen dominated. \\

In this work, we account for the wide variety of potential atmospheric compositions of hot sub-Neptunes and super-Earths. We calculate temperature-pressure profiles self-consistently from the gas phase composition, predict which condensates can be stable in these environments and  show the effect that clouds forming from these species can have on transmission spectra. In section $1$, we identify the most important condensates which can be stable in atmospheres of hot sub-Neptunes and super-Earths with equilibrium chemistry. In a second step we investigate the effect of the oxygen abundance of the atmosphere on the stability of these species on the example case of 55 Cnc e. We then compare the equilibrium chemistry results with a detailed cloud formation model. We simulate transmission spectra to verify whether different cloud species would be observable and distinguishable on a planet like 55 Cnc e. In section $4$, we use the cloud formation model to investigate the observability of clouds in \ce{O2}/\ce{CO2} atmospheres. Finally, we conclude in section $5$.

\section{Methods}

In this paper we focus on atmospheres of hot, sub-Neptune sized planets, including both sub-Neptunes and super-Earths. For simplicity, we refer to both populations as sub-Neptunes and use the planetary system 55 Cnc e as a representative case of both populations, since its radius and density situate the planet in between \citep{Hu2024}. Parameters are listed in Table \ref{Tab_55Cnce}, to set the conditions for the radiative transfer calculations. The goal is not to model this specific planet in detail, but to use it as an illustrative example for atmospheres of hot super-Earths and hot sub-Neptunes in general.

\begin{figure*}
    \centering
        \includegraphics[width=\textwidth]{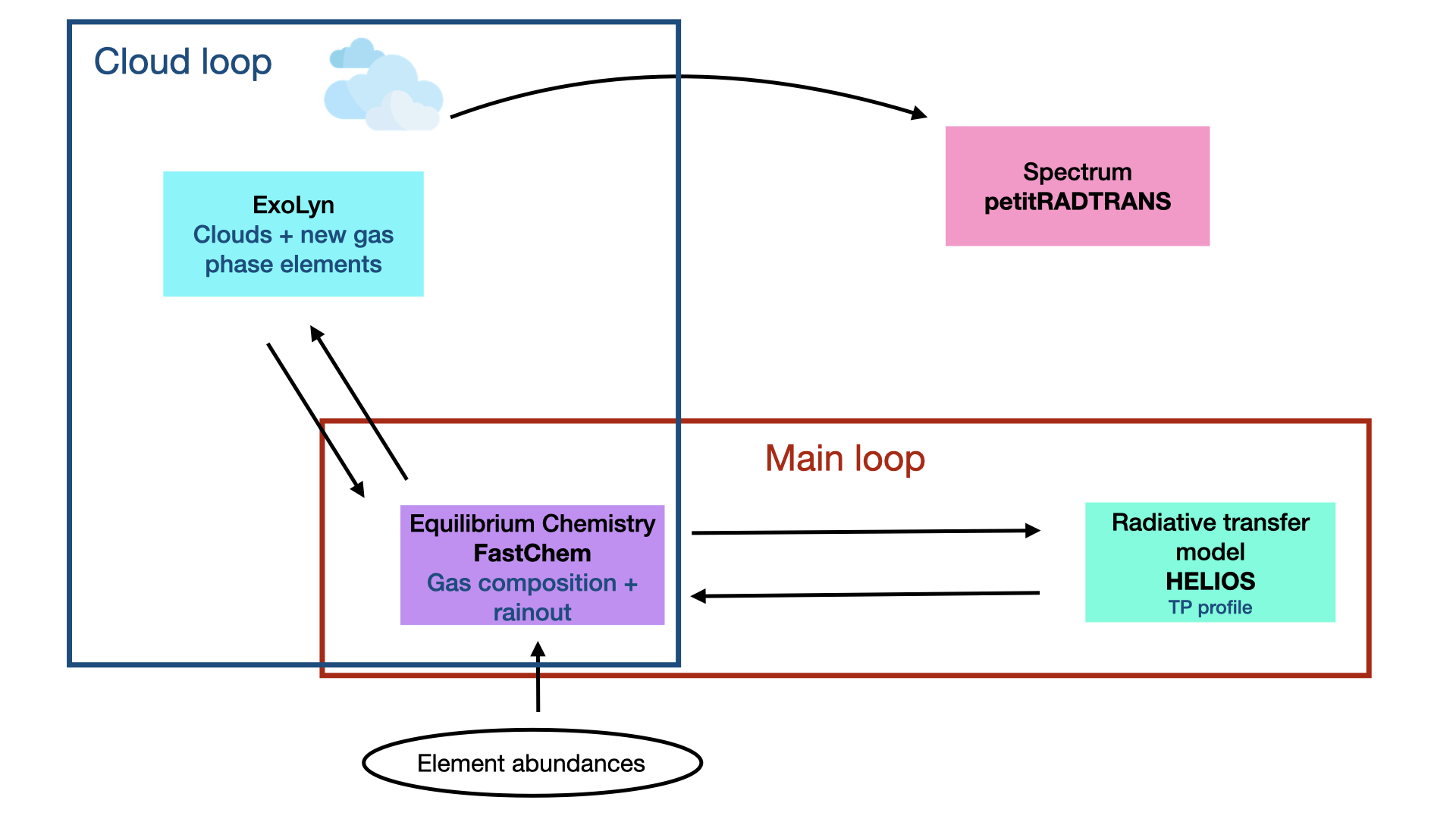}\\
    \caption{The sketch illustrates the different components of our pipeline. The main loop consists of self-consistent temperature profile and equilibrium chemistry computations achieved through iterations between {\textsc{HELIOS}} and {\textsc{FastChem}}3. Simple rainout is used in these computations to mimic the presence of a cloud in the atmosphere. For some selected cases we run the cloud loop, where we use the output of our main loop as input for the detailed cloud model {\textsc{ExoLyn}} and recompute the gas phase again with  {\textsc{FastChem}}3 after cloud formation. {\textsc{petitRADTRANS}} is called last to simulate transmission spectra from the outputs of the cloud loop.}
    \label{pipeline_sketch}
\end{figure*}

Our pipeline contains two separate loops, as illustrated in Figure \ref{pipeline_sketch}. The main loop is run for all our investigations and allows us to self-consistently compute the atmospheric temperature structure and equilibrium chemistry. The cloud loop and spectrum computation are only run for some selected models. With this part of the pipeline, we compute detailed cloud models to determine cloud composition, new gaseous abundances and spectral features. We explain the methodology in this section.

\subsection{Radiative transfer} \label{HELIOS}

\begin{table}
\centering
\caption{Star-planet parameters of the 55 Cnc e system}
\label{Tab_55Cnce}
\vspace*{-1mm}
\begin{tabular}{|c|c|} 
   \hline
   parameter & value\\ 
   \hline
   a & $0.016$\,AU \citep{Bourrier2018} \\
   Rp & $1.947$ R$_{\text{Earth}}$ \citep{Crida2018}\\
   Mp & $8.6$ R$_{\text{Earth}}$ \citep{Crida2018}\\
   Rs & $0.98$ R$_{\text{sun}}$ \citep{Crida2018}\\
   Ts &  $5214$\,K \citep{Hu2024}\\
   P$_{\text{surf}}$ &  $10$\,bar \citep{Hammond2017}\\ 
   \hline
\end{tabular}
\end{table}

We compute the temperature structure of the planet with {\textsc{HELIOS}} \citep{Malik2017,Malik2019a,Malik2019b,Whittaker2022}, accounting for absorption and scattering by species. {\textsc{HELIOS}} uses the two-stream approximation to calculate the radiative-convective equilibrium for a gas of a specified composition. We use our results from {\textsc{FastChem}}3 in combination with opacities from \citet{Zilinskas2023} with a resolution R$=2000$. We account for the opacities of following gaseous species: \ce{N2}, \ce{NaO}, \ce{NO}, \ce{SiN}, \ce{SO}, \ce{HNO3}, \ce{C2}, \ce{CaOH}, \ce{FeH}, \ce{H2+}, \ce{H3O+}, \ce{KOH}, \ce{OH+}, \ce{SiH2}, \ce{SiH4}, \ce{SiS}, \ce{N2O}, \ce{NaOH}, \ce{H2CO}, \ce{CH}, \ce{PC}, \ce{CS}, \ce{H2O2}, \ce{NH}, \ce{NS}, \ce{PH}, \ce{PO}, \ce{SO3}, \ce{SO2}, \ce{MgH}, \ce{AlH}, \ce{CaH}, \ce{TiH}, \ce{C2H2}, \ce{C2H4}, \ce{Mg}, \ce{SiO}, \ce{AlO}, \ce{CaO}, \ce{SiO2}, \ce{H2O}, \ce{CO}, \ce{CH4}, \ce{Al}, \ce{Ca}, \ce{Fe}, \ce{K}, \ce{MgO}, \ce{Na}, \ce{Si}, \ce{Ti}, \ce{TiO}, \ce{NH3}, NaH, SiH, \ce{H2S}, \ce{S}, \ce{PS}, \ce{CO2}, \ce{HCN}, \ce{OH}, \ce{CH3}, \ce{CN}, \ce{HS}, \ce{PH3}, \ce{O2}, \ce{H2}, H, He, bound free and free free emission of \ce{H-} and collision induced absorption of \ce{H2-H2}.
All line lists which we make use of are listed in Tables \ref{Tab_opacities} and \ref{Tab_opacities_2} with their references. We use a PHOENIX model \citep{Husser2013} for the stellar spectrum for the host star 55 Cnc A, which we generate with the star tool provided within {\textsc{HELIOS}}. In highly opaque regions, the atmosphere can become convective. In {\textsc{HELIOS}} this is accounted for by comparing the temperature gradient to a dry adiabat in each layer of the atmosphere. If the gradient is larger, it is reduced to an adiabat. For the purpose of computing the adiabatic temperature gradient in each atmospheric layer, we assume a diatomic atmosphere.

The rocky surface of the planet is taken at $10$\,bar. The upper limit of the atmosphere is set at $10^{-8}$\,bar. To take into account that the atmosphere redistributes some of the energy input from the star, we set the heat redistribution in {\textsc{HELIOS}} to $f=\frac{1}{3}$, like for the best fitting case of 55 Cnc e in \citet{Hu2024,vBuchem2024}. A value of $f=\frac{1}{4}$ corresponds to full heat distribution between day and night side, and $f=\frac{2}{3}$ corresponds to all heat being confined to the day side.

\subsection{Equilibrium chemistry with rainout}

To obtain a self-consistent atmospheric structure, the main loop of our setup is an iterative process between radiative transfer and chemistry models Figure \ref{pipeline_sketch}. We use {\textsc{HELIOS}} to obtain a temperature structure, then we run the chemistry code {\textsc{FastChem}}3 \citep{Stock2018,Kitzmann2024} to compute vertical mixing ratios in each atmospheric layer and feed these back to the radiative transfer model. We repeat this until convergence is reached

With the equilibrium chemistry code {\textsc{FastChem}}3 \citep{Stock2018,Kitzmann2024} we compute the abundances of $203$ gas species and ions and $109$ condensates in the atmosphere for a grid of compositions from element abundances of the $18$ elements \ce{H}, \ce{He}, \ce{N}, \ce{C}, \ce{P}, \ce{Cl}, \ce{S}, \ce{V}, \ce{O}, \ce{Na}, \ce{Mg}, \ce{Si}, \ce{Fe}, \ce{Al}, \ce{K}, \ce{Ca}, and \ce{Ti}.
The initial conditions of our pipeline are set by running {\textsc{FastChem}}3 on a wide grid of temperatures and pressures between $400-6000$\,K and $10^{-8}-1000$\,bar. In this grid, condensation is computed at each temperature-pressure point independently providing a first guess of the chemistry to obtain a temperature structure with {\textsc{HELIOS}}. This temperature structure is then used as a new input to {\textsc{FastChem}}3 to obtain vertical mixing ratios of all included species in the atmosphere. In this part of the iterative loop, we proceed with "rainout" modelling. This approach is optional in the {\textsc{FastChem}}3 code. It consists in condensation from high to low pressures, such that material condenses out layer by layer, where it is taken out of the atmosphere above. This changes the elemental budget at each altitude.

To help with convergence, we implement a gradual relaxation, by only allowing a $30\%$ of the predicted change in temperature structure from one iteration to the other. We consider a model as converged when the maximal temperature difference between two iterations is on the order of $10$\,K. This relaxation method prevents the model from oscillating between hot, cloud-free cases and colder, cloudy cases.

\subsection{The cloud model} \label{m_exolyn}

Equilibrium chemistry provides an indication on the stability of solid and liquid species and can be used as a first guess for the extent and the location of cloud formation. However, disequilibrium processes are crucial to include for two major reasons: \\

1) Rainout chemistry underestimates the cloud mass, because at each atmospheric layer material is taken out of the atmosphere when it rains out. No replenishment from vertical movement in the atmosphere is taking place without including dynamics.\\

2) Equilibrium chemistry does not provide an estimate of the cloud particle size, but only the condensate abundance. To form cloud particles, nucleation, coagulation, vertical transport, and settling are required in addition to condensation and evaporation processes. \\

Hence, for some selected cases we use a detailed cloud model for comparison, running he cloud loop  Figure \ref{pipeline_sketch} after completion of the main loop. We predict cloud formation in atmospheres of hot, rocky sub-Neptunes with {\textsc{ExoLyn}} \citep{Huang2024}. {\textsc{ExoLyn}} solves vertical transport for gas and solid species, computes condensation from a pure gas phase and takes into account coagulation and nucleation to compute a vertical distribution of the cloud's mass mixing ratio and grain sizes. The model takes as input the gas phase abundance of all cloud forming species at the bottom of the atmosphere, a temperature-pressure structure, an eddy diffusion parameter $\text{K}_{\text{zz}}$, a particle diffusion parameter $\text{K}_{\text{p}}$, nucleation parameters to compute the nucleation profile according to \citet{Ormel2019}, and cloud forming reactions. {\textsc{ExoLyn}} follows the Bruggemann model \citep{Huang2024,Kiefer2024}, assuming one particle size per cloud layer. In the Bruggemann model, each cloud particle is a mixture of all condensing material in the layer.

We use the temperature structure from our radiative transfer calculations with {\textsc{HELIOS}} and run {\textsc{FastChem}} with rainout to compute the gas phase abundances at the bottom of the atmosphere. Rainout is implemented here to account for the material which remains at the surface in solid or liquid form.  We keep nucleation parameters as in \citet{Huang2024} shown in Table \ref{Tab_nuc_parms}. Nucleation is parameterised in {\textsc{ExoLyn}} with a lognormal nucleation profile as in \citet{Ormel2019}. We note that the chosen values for nuclei production rate, width of the nucleation profile and nucleation height affect the extent of the cloud and the particle size. However, these parameters are very uncertain and are dependent on the cloud species and on the temperature \citep[e.g.][]{Gao2020}. In addition, \citet{Huang2024} find that the nucleation profile is not the dominating factor determining the cloud profile. Therefore, we keep their default values. We set our default value for eddy diffusion for gas and solid particles to $\text{K}_{\text{zz}}=10^{10}$ cm$^2$s$^{-1}$ and $\text{K}_{\text{p}}=10^{10}$ cm$^2$s$^{-1}$. For the mean molecular weight of the atmosphere $m_\mathrm{gas}$, we take an average of the mean molecular weight over the entire pressure range. 

\begin{table}
\centering
\caption{Cloud formation parameters }
\label{Tab_nuc_parms}
\vspace*{-1mm}
\begin{tabular}{|c|c|} 
   \hline
   parameter & value\\ 
   \hline
   $\sigma_\text{mol}$ (molecular cross section) &  $2\cdot10^{-15}$ [cm$^2$]\\
   f$_{\text{stick}}$ (sticking probability)& $1$\\
   f$_{\text{coag}}$ (coagulation probability)& 1\\
   $\sigma_{\text{com}}$ combined cross section for vapour and \ce{H2} & $8\cdot10^{-15}$ [cm$^2$] \\
   $\rho_{\text{int}}$ particle internal density& $2.8$ [g cm$^{-3}$]\\ 
   $\text{a}_{\text{p0}}$ (nucleation radius)&  $10^{-7}$ [cm]\\
   $\textbf{m}_{n0}$ nucleation mass&  $\frac{4}{3}\pi\cdot\rho_{int} \cdot \text{a}_{p0}^3$ [g]\\
   $\dot{\Sigma_n}$ (nuclei production rate)& $10^{-15}$ [g cm$^{-2}$ s$^{-1}$] \\
   $\sigma_\text{n}$ (width nucleation profile)& $0.2$ \\
   P$_{\text{n}}$ (nucleation height) &  $60$ [g cm$^{-2}$ s$^{-1}$] \\
   \hline
\end{tabular}
\end{table}

{\textsc{ExoLyn}} takes as input only one reaction per condensate. It is crucial to be careful when picking the reaction rates for condensate formation because the stability of a condensate is strongly dependent on the reaction rate and the amount of available elements to condense.
For the condensates for which reaction rates are available in the literature we have considered reaction rates from \citet{Helling&Woitke2006,Helling2008,Helling2017,Huang2024} and settled on the optimal choice resulting in the highest condensate mass. For the species for which condensation reactions are not available from literature, we have evaluated different reactions with combinations of reactants from the most abundant gas-phase species containing the elements required to form the condensate and have also settled on the optimal choice. The resulting condensation reactions which we consider are listed in Table \ref{Tab_reactions}.

\begin{table*}
\begin{center}
\caption{Condensation reactions $^{(1)}$ }
\label{Tab_reactions}
\vspace*{-1mm}  
\begin{tabular}{|c|c|} 
   \hline
   Cloud species & Formation reaction \\ 
   \hline
   & \ce{O}-dominated atmosphere \\
   \ce{SiO2(s)}& \ce{SiO + H2O -> SiO2(s) + H2}\\
   \ce{TiO2(s)}  &\ce{TiO2(g) -> TiO2(s)}\\
   \ce{MgTi2O5(s)} & \ce{Mg(OH)2 + 2TiO2 -> MgTi2O5(s) + H2O} \\
   \ce{Mg2TiO4(s)} & \ce{Mg(OH)2 + TiO2 -> Mg2TiO4(s) + 2H2O} \\
   \ce{MgSiO3(s)} & \ce{Mg(OH)2 + SiO -> MgSiO3(s) + H2} \\
   \ce{Mg2SiO4(s)} & \ce{2Mg(OH)2 + SiO -> Mg2SiO4(s) + H2O + H2} \\
   \ce{MgAl2O4(s)} & \ce{Mg(OH)2 + 2HAlO2 -> MgAl2O4(s) + 2H2O} \\
   \ce{Fe2O3(s)} & \ce{2FeO + H2O -> Fe2O3(s) + H2}\\
   \ce{Fe3O4(s)} & \ce{3FeO + H2O -> Fe3O4(s) + H2}\\
   \ce{FeO(s)} & \ce{FeO -> Fe3O4(s)}\\
   \ce{Fe2SiO4(s)} & \ce{2FeO + SiO + H2O -> Fe2SiO4(s) + H2}\\
   \ce{CaSiO3(s)} & \ce{Ca(OH)2 + SiO -> CaSiO3(s) + H2}\\
   \ce{CaTiO3(s)} & \ce{Ca(OH)2 + TiO2 -> CaTiO3(s) + H2O}\\
   \ce{Mg3(PO4)2(s)} & \ce{3Mg(OH)2 + 2PO2 -> Mg3(PO4)2(s)(s) + 2H2O + H2}\\
   \ce{CaMg(SiO3)2(s)}& \ce{Ca(OH)2 + Mg(OH)2 + 2SiO -> CaMg(SiO3)2(s) + 2H2}\\   
   \hline
    &\ce{C}-dominated atmosphere\\
    \ce{Fe(s)} & \ce{Fe -> Fe(s)}\\
    \ce{Al2O3(s)} & \ce{2Al + 3H2O -> Al2O3(s) + 3H2}\\
    \ce{C(s)} & \ce{C2H2 -> 2C(s) + H2}\\
    \ce{TiC(s)} & \ce{CH4 + TiO -> TiC(s) + + H2O + H2}\\
   \hline
\end{tabular}\\
\end{center}
$^{(1)}$ {\small Condensation reactions for each solid species in our cloud formation model. These pathways are chosen according to the availability of reactants and their efficiency to produce condensates.}
\end{table*}

\subsection{Cloud features in transmission spectroscopy} \label{pRT}

To assess whether potential clouds would affect observations of hot sub-Neptunes, we first of all investigate potential features of the cloud species for which we have optical data. We then simulate transmission spectra for some selected cases between $0.1-30$\,$\mu$m. We collect the refractive indices of the solid species which appear in our models and are available from either the JENA database of optical constants\footnote{\url{https://www.astro.uni-jena.de/Laboratory/OCDB/index.html}} or already implemented into \textsc{ExoLyn} from \citet{Kitzmann&Heng2018}. Table \ref{opconst} lists all the references for the opacities of solid species relevant for this work. We then use {\textsc{optool}} \citep{Dominik2021} to predict extinction coefficients for each of these species. Cloud particles are not expected to be perfectly spherical and their geometry affects scattering and absorption properties. Therefore, we assume a distribution of hollow spheres (DHS) to generate the cloud opacities. For solid particle agglomerations like clouds, DHS is a good approximation to model grain-light interactions \citep{Min2005}.

\begin{table}
\centering
\caption{\textbf{References} for optical constants of solid species}
\label{opconst}
\vspace*{-1mm}
\begin{tabular}{|c|c|} 
   \hline
   solid material & source for optical data\\ 
   \hline
   \ce{Al2O3(s)} & \citet{Koike1995}\\
   \ce{C(s)} & \citet{Draine2003}\\
   \ce{CaTiO3(s)} & \citet{Palik1991}\\
   \ce{Fe(s)} & \citet{Palik1991}\\
   \ce{Fe2O3(s)} & A.H.M.J. Triaud \\
   \ce{Fe2SiO4(s)} & \citet{Fabian2001}\\
   \ce{FeO(s)} & \citet{Henning1995}\\
   \ce{FeS(s)} & \citet{Pollack1994}\\
   \ce{Mg2SiO4(s)} & \citet{Jaeger2003}\\
   \ce{MgSiO3(s)} & \citet{Jaeger2003}\\
   \ce{MgAl2O4(s)} & \citet{Jaeger2003}\\
   \ce{MgO(s)} & \citet{Palik1991}\\
   \ce{SiC(s)} & \citet{Laor1993}\\
   \ce{SiO(s)} & \citet{Palik1985}\\
   \ce{SiO2(s)} & \citet{Palik1985,Henning1997}\\
   & \citet{Zeidler2013}\\
   \ce{TiC(s)} & \citet{Henning2001}\\
   \ce{TiO2(s)} & \citet{Posch2003,Zeidler2011}\\
   \hline
\end{tabular}
\end{table}

In a first part, we predict extinction coefficients for each solid species from Table \ref{opconst} between $0.1-1000$\,$\mu$m and model the extinction of small particles with a diameter of $0.01$\,$\mu$m composed of one single material. In a second part, we study the effect of cloud opacities for the same selected cases which we choose for the cloud study described in section \label{m_exolyn}. For this investigation we run the entire cloud loop, including rainout within the equilibrium chemistry model and transmission spectra as illustrated in Figure \ref{pipeline_sketch}. Rainout is important to account for in this case because some material which is liquid or solid already at the magma ocean pressure would rain out immediately without ever reaching the cloud formation layer. After the cloud computation with {\textsc{ExoLyn}}, we recompute the atmospheric gas chemistry with {\textsc{FastChem}}3 from the remaining elements to get the mixing ratios of major gas phase molecules. 

We generate cloud opacities with {\textsc{optool}} from grain sizes computed with {\textsc{ExoLyn}} and the relative mass fractions of the different components in the cloud particles. We run petitRADTRANS \citep{Molliere2019} to predict transmission and emission features of the cloudy atmosphere, investigating a pressure range of the atmosphere from $10$-$10^{-8}$\,bar with the temperature structures from our radiative transfer computations. We use gas phase line by line opacities for the spectra modelling in petitRADTRANS from the ExoMol data base \citep{Tennyson2024} at resolution $\lambda/\Delta\lambda = 1000$ for the following species: \ce{Ti+}, \ce{TiO}, \ce{TiH}, \ce{Ti}, \ce{SO}, \ce{SO2}, \ce{SO}, \ce{SiS}, \ce{SiO2}, \ce{SiO}, \ce{SiH4}, \ce{SiH2}, \ce{SiH}, \ce{Si}, \ce{SH}, \ce{PS}, \ce{PO}, \ce{PN}, \ce{PH3}, \ce{PH}, \ce{OH}, \ce{COS}, \ce{O3}, \ce{O2}, \ce{O}, \ce{NO}, \ce{NH3}, \ce{NH}, \ce{NaOH}, \ce{NaH}, \ce{Na}, \ce{N2O}, \ce{MgO}, \ce{MgH}, \ce{Mg}, \ce{K}, \ce{HNO3}, \ce{HCN}, \ce{H3O+}, \ce{H2S}, \ce{H2CO}, \ce{H2}, \ce{FeH}, \ce{Fe}, \ce{Fe+}, \ce{CP}, \ce{CO2}, \ce{CO}, \ce{CN}, \ce{CH4}, \ce{CaOH}, \ce{CaO}, \ce{CaH}, \ce{C2H4}, \ce{C2H2}, \ce{C2}, \ce{AlO}, \ce{AlH}, \ce{AlF}, \ce{Al}, \ce{N2}, \ce{He}, \ce{e-}, \ce{H-} and \ce{H}. In addition we include rayleigh species \ce{H2} and He as well as continuum opacities from \ce{H2}-\ce{H2}, \ce{H2O}-\ce{H2O}, \ce{N2}-\ce{O2}, \ce{N2}-\ce{N2}, \ce{O2}-\ce{O2}, \ce{N2}-He and \ce{H2O}-\ce{N2}. For the line lists used and there references see Table \ref{Tab_opacities} and Table \ref{Tab_opacities_2}.

\subsection{The grid of compositions} \label{grid_methods}

During planet formation, volatiles get trapped into a rocky planet's interior. Due to outgassing, secondary atmospheres composed of these volatiles as well as more heavy elements such as \ce{Si}, \ce{Fe}, \ce{Mg}, \ce{Na} and \ce{K}  can build up around the planet \citep[e.g.][]{vBuchem2023, vBuchem2024,Cherubim2025,Lichtenberg2025}. Some of the sub-Neptune sized planets might also have a primary hydrogen envelope into which the outgassed material is mixed \citep[e.g.][]{vBuchem2024,Cherubim2025}. Therefore, the expected compositions span a wide variety. Our aim in this work is to cover such a variety in atmospheric compositions, focusing on how different ratios of volatiles would impact condensation and cloud formation. For this purpose, we construct a grid of models with $144$ different sets of element abundances. 

We vary the ratios of the volatile species carbon (\ce{C}), hydrogen (\ce{H}), nitrogen (\ce{N}), oxygen (\ce{O}) , phosphorus (\ce{P}) and sulphur (\ce{S}) while keeping all other elements in solar abundances relative to each other and relative to the total volatile content. We take these solar values from \citet{Lodders2019}. We divide our models into three types of atmospheres: \ce{N}-dominated atmospheres, atmospheres that are either \ce{C}- \ce{CO}- or \ce{O}-dominated, and \ce{H}-dominated atmospheres. This choice is motivated by the different atmospheric scenarios proposed by \citet{Hu2024,Zilinskas2020,Zilinskas2023}. From these 3 atmospheric types we create a set of 9 (3x3) models, for which the fraction of the dominating volatile $X_H$, $X_C+X_O$ or $X_N$, varies from 60\% to 75\% to 90\%. We explore \ce{C}/\ce{O} ratios $0.1$, $0.5$, $1$ and $2$, expanding this grid to a 3x3x4 grid of models. We note that these \ce{C}/\ce{O} ratios are taken regarding the bulk composition. Condensation and cloud formation can alter the \ce{C}/\ce{O} ratio of the gas phase. The carbon- and oxygen- dominated atmospheres are \ce{O}-atmospheres if \ce{C}/\ce{O}$<1$,  \ce{CO}-atmospheres if \ce{C}/\ce{O}=1 and \ce{C}-atmospheres if \ce{C}/\ce{O}$>1$.

The sulphur and phosphorus content of rocky planets can also be expected to vary drastically due to different sulphur and phosphorus budgets in stars \citep{Hinkel2014}, formation mechanisms and processes like outgassing and volcanism which can enhance the abundance of these volatiles as it is the case on Io \citep{Lellouch2007,Gaillard2014,vBuchem2024}.  In order to account for these various conditions, we explore a sulphur and phosphorus rich case (\ce{P}+\ce{S} rich) and a sulphur and phosphorus poor case (\ce{P}+\ce{S} poor) for each of the models in the 3x3x4 grid above, extending it to a 3x3x4x2 grid with 72 models. The \ce{P}+\ce{S} poor cases have abundances $X_S=X_P=0.1$\,ppm and the \ce{P}+\ce{S} rich cases have enhanced sulphur and phosphorus fractions of $X_S=X_P=1$\,\%. For reference, solar abundances are $X_S=1.7\cdot10^{-5}$ and $X_P=3.3\cdot10^{-7}$ \citep{Lodders2019}.
So far, we have attributed the sulphur and phosphorus abundance as well as the abundances of the major volatile. For each model to include all species and fill up to a fraction of unity, we attribute the leftover fraction to the two leftover volatiles in  a systematic way. The smallest volatile fraction among $X_H$, $X_C+X_O$ and $X_N$ is set to $X_\epsilon=0.01$. The last leftover species is the assigned an abundance such that  the total fraction $X_H+X_C+X_O+X_C+X_O+X_N=1$.
To summarise, we have assigned the abundance of the major volatile, the sulphur abundance and the phosphorus abundance. The least abundant volatile in the grid except for sulphur and phosphorus is assigned a fraction of $X_\epsilon = 0.01$. Since the total of $X_H+X_C+X_O+X_C+X_O+X_N=1$, we then determine the elemental abundance of the remaining volatile - or volatiles in the cases where it is \ce{C}+\ce{O} which we are missing - by filling up the total fraction to unity.

\subsection{How oxygen abundance affects condensation}\label{redox}

The grid study presented above covers a wide range of atmospheric oxygen abundances. In a second exploration, we focus on the effect that the oxygen abundance has on condensation. For this purpose we focus on one particular example, which is the carbon richer model of 55 Cnc e from \citet{vBuchem2024}.It is to note, that this composition is not carbon rich, but \citet{vBuchem2024} refer to it as the carbon rich case since they explore another case as well which is even poorer in carbon.\\ We introduce this new composition in this section for a more reliable example study. The compositions explored in previous section were a purely explorative parameter study. However, in order to make predictions which are relevant for potentially observable exoplanet compositions, we now base our more detailed investigations on self-consistent models which were selected from fits to data of JWST observations by \citet{vBuchem2024,Hu2024}.

\begin{figure*}
    \centering
          \includegraphics[width=\textwidth]{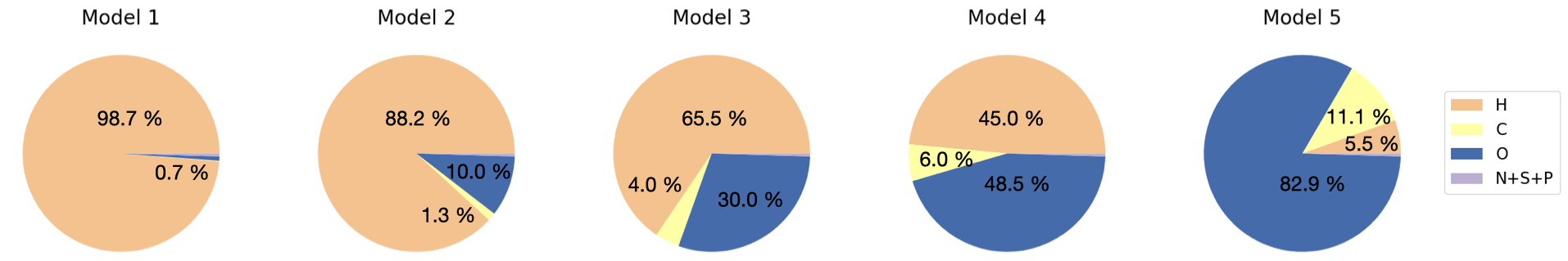}
        \caption{The pie charts show the percentage of the five  volatiles in the atmosphere of the 55 Cnc e-type atmosphere and the four carbon- and oxygen- enriched cases. Orange represents the mole fraction of atomic hydrogen, blue the oxygen fraction and yellow shows the carbon fraction. In all five models, \ce{C}/\ce{O}$=0.133$ and the mole fractions of nitrogen, sulphur and phosphorus are kept constant with \ce{N}$=0.0008012$, \ce{S}$= 0.002597867$ and \ce{P}$=0.0013388$. Their sum is indicated in purple in the pie charts.}
\label{fig_piecharts}
\end{figure*}

The atmosphere by \citet{vBuchem2024} is \ce{H}-dominated with a \ce{C}/\ce{O} ratio of $0.133$. We then further consider four carbon- and oxygen- enriched cases of this atmosphere by increasing the \ce{O}/\ce{H} ratio, wile keeping the \ce{C}/\ce{O} ratio and the \ce{N}, \ce{P} and \ce{S}- abundances constant. Figure \ref{fig_piecharts} illustrates how this changes the elemental composition of the atmosphere for each of the 5 cases. All bulk refractory abundances are taken to be solar relative to each other and relative to the total volatile content, similar to our grid study described in section \ref{grid_methods}.

\section{Results}

In this section we explore which cloud species we can expect to find in atmospheres of hot sub-Neptunes using the main loop of our pipeline Figure \ref{pipeline_sketch}. For some selected cases we investigate how these predictions compare to a more detailed cloud model which we run with the cloud loop Figure \ref{pipeline_sketch}. In these atmospheres we search for characteristic absorption features generated by pure and mixed cloud grains.

\subsection{Expected condensates in hot atmospheres of sub-Neptunes}\label{condensates}

In this first part, we show which cloud species we can expect to find in the atmospheres of hot sub-Neptunes depending on their composition. Note, that here we focus on the major condensates which could be stable in different types of atmospheres without studying them in detail and without incorporating a complex cloud model. We provide an overview of the condensates which can form in carbon (\ce{C})- oxygen (\ce{O})- nitrogen (\ce{N})-dominated and hydrogen (\ce{H})-dominated atmospheres. We also identify patterns in the temperature structures in the different atmospheric types. 

\begin{figure}
\centering
        \includegraphics[width=0.49\textwidth]{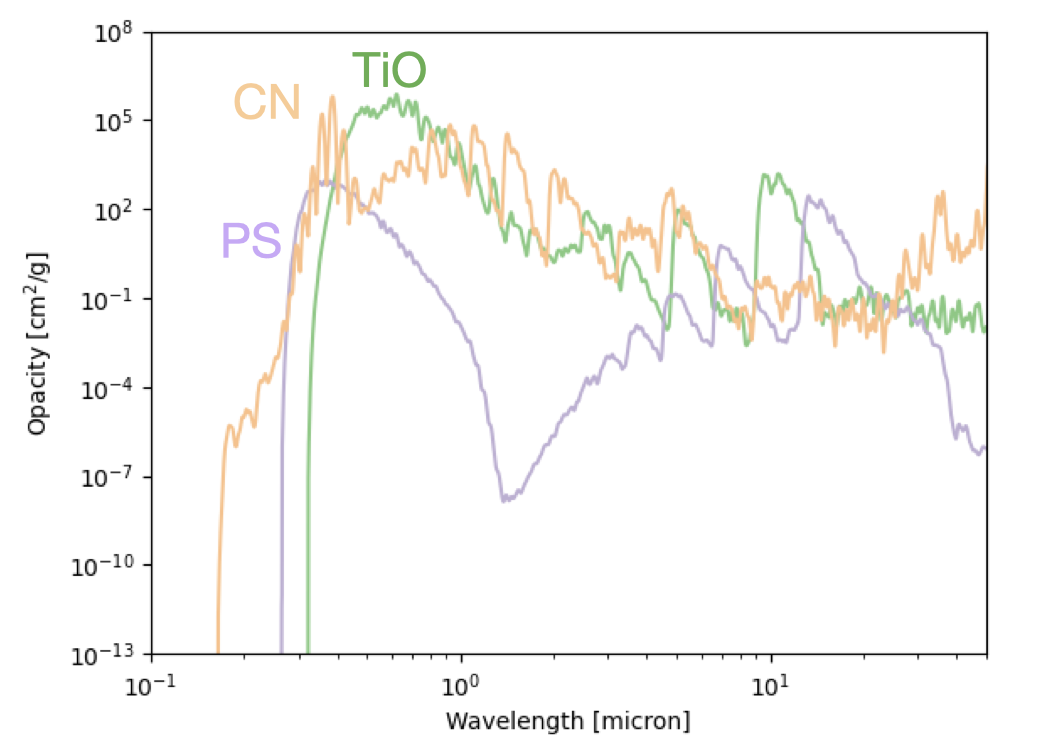} %
\caption{Opacities of some relevant strongly UV and optical absorbing species at $2000$\,K and $10^{-5}$\, bar. The altitude is chosen according to where the opacity contribution is the strongest.}
    \label{opacities}
\end{figure}

Figure \ref{fig:grid_TPs} shows an overview of the temperature-pressure structures of our grid of atmospheres. Each panel corresponds to twelve atmospheres for one atmospheric type with a certain \ce{C}/\ce{O} ratio. The rows show the three types from top to bottom: Type 1: \ce{O}-, \ce{CO}-, or \ce{C}-dominated (depending on the \ce{C}/\ce{O} ratio), type 2: \ce{N}-dominated and type 3: \ce{H}-dominated. The columns indicate the \ce{C}/\ce{O} ratio of the atmosphere from left to right with \ce{C}/\ce{O}=0.1, \ce{C}/\ce{O}=0.5, \ce{C}/\ce{O}=1 and \ce{C}/\ce{O}=2. The temperature profiles are split into two categories. Blue temperature profiles correspond to atmospheres with low elemental sulphur and phosphorus abundance, $X_{\text{S}}= X_{\text{P}} =0.1$\,ppm and red temperature profiles show models with high elemental sulphur and phosphorus abundance of $X_{\text{S}}= X_{\text{P}} =1$\,\%. This split illustrates the behaviour that sulphur and phosphorus rich atmospheres are hotter if the atmosphere is oxygen poor and hydrogen rich. \ce{PS} can form in these cases and create an inversion in the atmosphere due to strong absorption in the UV and optical, see Figure \ref{opacities}. We focus on condensates which reach abundances of $>1$\,ppb. These are illustrated in Figure \ref{fig:grid}, which is organised in the same format as Figure \ref{fig:grid_TPs}. A choice of refractory abundances larger than solar could shift the results towards more stable condensate species and higher abundances. It is also to note that we don't take into account the altitude of formation and the extent of a condensate to determine its importance.\\

\subsubsection{\ce{O}-, \ce{CO}-, and \ce{C}-dominated atmospheres}
In \ce{O}-dominated atmospheres (\ce{C}/\ce{O}$<1$) the near-surface region is the hottest part of the atmosphere because of the presence of greenhouse gases like \ce{H2O} and \ce{CO2}. The temperature decreases with altitude reaching the lowest values of all our models with T$<1000$\,K at P$<10^{-7}$\,bar. This behaviour allows for a large variety of condensates to be stable. Due to the temperature structures being so similar for all the models in this atmospheric type, many of the condensates can be stable in multiple of these atmospheres. This is illustrated by Figure \ref{fig:grid}, where a large fractional occurence is reached for multiple species in these atmospheres.

At \ce{C}/\ce{O}=0.1 we can expect the following condensates: \ce{TiO2(s,l)}, \ce{SiO2(s,l)}, \ce{MgTi2O5(s,l)}, \ce{MgSiO3(s,l)}, \ce{MgAl2O4(s,l)}, \ce{Mg3(PO4)2(s,l)}, \ce{Mg2TiO4(s,l)}, \ce{Fe3O4(s,l)}, \ce{Fe2O3(s,l)}, \ce{CaTiO3(s)}, \ce{CaSiO3(s)}, \ce{CaMgSi2O6(s)}, \ce{Al6Si2O3(s,l)}, \ce{Al2O3(s,l)} and \ce{(P2O5)2(s)}. Out of these species, \ce{SiO2(s,l)}, \ce{TiO2(s,l)}, \ce{CaTiO3(s)} and \ce{Fe3O4(s,l)} are stable in all atmospheres of this type.

If the \ce{C}/\ce{O} ratio is increased to \ce{C}/\ce{O}=0.5, \ce{(P2O5)2(s)}, \ce{Al6Si2O3(s,l)}, \ce{CaSiO3(s)} and \ce{Fe2O3(s,l)} do not form in our models anymore, but instead we find \ce{Mg2SiO4(s,l)}, \ce{Fe2SiO4(s)}, \ce{FeO(s,l)} and \ce{MgTiO3(s,l)}. Similarly to the case with \ce{C}/\ce{O}=0.1, \ce{CaTiO3(s)}, \ce{SiO2(s,l)}, and \ce{TiO2(s,l)} are also stable in all models of this kind, in addition to \ce{Fe2SiO4(s)} and \ce{MgAl2O4(s,l)}. 

These differences show, that iron species are particularly sensitive to changes in the \ce{C}/\ce{O} ratio in \ce{O}-atmospheres. At \ce{C}/\ce{O}=0.1 iron condenses mainly in it's most oxidised forms \ce{Fe3O4(s)} and \ce{Fe2O3(s)}. This changes as the \ce{C}/\ce{O} ratio is increased to 0.5 where major iron condensates are \ce{Fe2SiO4(s)} and \ce{FeO(s,l)}. However,  aluminium condensates  and magnesium species are also affected since \ce{Al2O3(s,l)} forms frequently in addition to \ce{MgAl2O4(s,l)} at \ce{C}/\ce{O}=0.5 only and magnesium-oxides are overall less frequently found at \ce{C}/\ce{O}=0.1.

In \ce{CO}-atmospheres (carbon and oxygen dominated atmospheres with (\ce{C}/\ce{O}=1), the temperatures reached at low pressures are much higher than in carbon poorer models. Only half of the \ce{CO}-atmospheres reach temperatures $<1250$\,K, whereas all models cool down below this threshold if \ce{C}/\ce{O}$<1$. The gas phase of \ce{CO}-atmospheres is dominated by the molecule \ce{CO}, because carbon and oxygen have similar bulk abundances. Fewer solid and liquid species can be stable in fewer models compared to \ce{O}-atmospheres, see Figure \ref{fig:grid}. The most common condensates in \ce{CO}-atmospheres are \ce{Al2O3(s,l)}, \ce{CaTiO3(s)}, and \ce{TiO2(s,l)} condensing in $75\%$, $60\%$ and $60\%$ of the models respectively. In addition, we find \ce{MgSiO3(s,l)}, \ce{SiO2(s,l)}, \ce{MgTi2O5(s,l)}, \ce{Mg2SiO4(s,l)}, \ce{Fe(s,l)}, \ce{Ti2O3(s,l)} and \ce{TiN(s,l)}. In the four coldest atmospheres condensation sets on at altitudes between $5 \cdot 10^{-3} $\,bar and $10^{-4}$\,bar. \ce{Al2O3(s,l)}, \ce{CaTiO3(s)} and \ce{TiO2(s,l)} are always the species condensing first. In the \ce{P}+\ce{S} poor \ce{CO}-atmospheres with the highest \ce{H}/\ce{O} ratio as well as in the \ce{P}+\ce{S} rich \ce{CO}-atmospheres with the lowest \ce{H}/\ce{O} ratio (the two turquoise and the dark red model), condensation can only happen between $10^{-4}-10^{-7}$\,bar. Mostly \ce{TiO2(s,l)}, \ce{CaTiO3(s)} and \ce{Al2O3(s,l)} condense in these regions. The medium and dark orange temperature profiles correspond to the atmospheres which are \ce{P}+\ce{S} poor, have a high \ce{H}/\ce{O} ratio and contain more nitrogen compared to hydrogen. This means that the bulk oxygen abundance in these cases is particularly low while nitrogen and carbon are abundant and the temperatures are low enough, resulting in the stability of very reducing species like \ce{TiC(s)} and \ce{TiN(s)}. In the case which is \ce{P}+\ce{S} rich and hydrogen is the second most abundant volatile after carbon, the conditions are such that condensation cannot occur at all: The \ce{H}/\ce{O} ratio is high and \ce{PS} can be stable, leading to temperatures which are high throughout the entire atmospheres.

\ce{C}+\ce{O}-dominated atmospheres with \ce{C}/\ce{O}=2 are \ce{C}-dominated in their bulk composition. Despite these high total \ce{C}/\ce{O} ratio, the gas phase \ce{C}/\ce{O} often decrease to approximately unity. We will address this at the end of this section. Out of these models, the six compositions with low \ce{P}+\ce{S} abundance all show a decrease in temperature with altitude. The less carbon and oxygen and the more hydrogen, the hotter the atmosphere becomes at low pressures. The six \ce{C}-dominated models rich in \ce{P}+\ce{S} have a temperature inversion and no condensation. The most reducing cases with more carbon and oxygen and less hydrogen (red models) cool with decreasing pressure above the inversion. Three condensation behaviours show depending on the temperature structures: both, oxides and carbides condense, only carbides condense or the atmosphere is purely gaseous. As shown in Figure \ref{fig:grid}, we find predominantly \ce{TiC(s)}, \ce{C(s)} and \ce{Al2O3(s,l)} in respectively $60\%$, $50 \%$ and  $40\%$ of our \ce{C}-dominated models. Condensates can form between $1$-$10^{-2}$\,bar where there is a strong temperature drop. The four coldest models (blue) reach low enough temperatures for \ce{C(s)}, \ce{Al2O3(s,l)}, \ce{MgO(s)}, \ce{(CaAl)2SiO7(s)},  \ce{CaSiO3(s)}, \ce{Ca2SiO4(s)}, \ce{SiO(s)}, \ce{Fe(s,l)}, \ce{Mg2SiO4(s,l)} and \ce{MgSiO3(s,l)} to form. Two condensate layers of \ce{TiC(s)} and \ce{C(s)} can form in the two light turquoise models where the temperature drops to $1250$\,K - $1300$\,K at $10^{-8}$\,bar, enhancing the occurrence rate of these two species in Figure \ref{fig:grid} further.

We find that the \ce{C}/\ce{O} ratio of the gas phase in the \ce{C}-atmospheres is only around 10\% above unity. This is a consequence of heavy graphite rainout to the surface in these atmospheres. Hence, gaseous \ce{CO} is the major carbon and oxygen carrier there. Temperature drops with altitude can lead to the stability of some oxygen bearing condensates in these atmospheres. At low enough temperatures and high enough pressures, oxygen can still be thermally stable in liquid and solid compounds even if the \ce{C}/\ce{O} ratio is close to unity. In certain cases, carbides and - in nitrogen-rich atmospheres - nitrides can rain out at low altitudes. This decreases the \ce{C}/\ce{O} ratio further with increasing altitude. \\

\subsubsection{\ce{N}-dominated atmospheres}
In \ce{N}-dominated atmospheres, there is a bifurcation in the temperature structures at each \ce{C}/\ce{O} ratio, see Figure \ref{fig:grid_TPs}. If \ce{C}/\ce{O}=0.1 or \ce{C}/\ce{O}=0.5, temperature profiles of oxygen poorer atmospheres with high \ce{H}/\ce{O} ratio (turquoise and orange models) are inverted. \ce{TiO} can create these inversions around  $10^{-4}$\,bar  (turquoise models). If \ce{PS} forms as additional heating agent (orange models) it can lower the altitude of the inversions to pressures $>10^{-3}$\,bar. In models with low \ce{H}/\ce{O} ratio (dark blue and red), sulphur and phosphorus mostly form \ce{PO2} and \ce{SO2} and titanium is mostly in the form of \ce{TiO2}. Hence, the surfaces are hot and the atmospheres cool with altitude. 

The only cases for which condensation is possible, are the atmospheres with low \ce{H}/\ce{O}. Thus solids and liquids form in fewer models compared to the \ce{O}-dominated atmospheres, see Figure \ref{fig:grid}. At \ce{C}/\ce{O}$<1$, the most common condensates appear in $60\%$ of the \ce{N}-atmospheres. \ce{TiO2(s,l)}, \ce{SiO2(s,l)}, \ce{MgTi2O5(s,l)}, \ce{MgAl2O4(s,l)}, \ce{Mg3(PO4)2(s,l)}, \ce{Fe3O4(s,l)}, \ce{CaTiO3(s)}, and \ce{Al2O3(s,l)} form at both  \ce{C}/\ce{O}=0.1 and \ce{C}/\ce{O}=0.5. Along those we find \ce{Mg2TiO4(s,l)}, \ce{CaMgSi2O6(s)}, \ce{Al6Si2O3(s,l)}, and \ce{(P2O5)2(s)} at \ce{C}/\ce{O}=0.1, and \ce{Mg2SiO4(s,l)}, \ce{MgSiO3(s,l)}, \ce{Fe2SiO4(s)} and \ce{FeO(s,l)}  at \ce{C}/\ce{O}=0.5.

In \ce{N}-dominated atmospheres with \ce{C}/\ce{O}=1, the thermal structures are majorly affected by the \ce{P}+\ce{S} content, rather than the \ce{H}/\ce{O} ratio. The red and orange models are hotter than the blue models and all have inversions between $0.1-10^{-3}$\,bar. The \ce{P}+\ce{S} poor models (blue and turquoise) have non existing to small inversions at high pressures. The models with higher \ce{H}/\ce{O} ratios have inversions at low pressures created by atomic \ce{Fe}. The only three cases where condensation can proceed are the dark blue models with low \ce{H}/\ce{O} ratio and low \ce{P}+\ce{S} abundance. The two stable condensates forming there are \ce{TiN(s,l)} and \ce{Al2O3(s,l)}.

\ce{N}-dominated atmospheres with \ce{C}/\ce{O}=2 show three types of temperature structures:
The atmospheres with low \ce{P}+\ce{S} abundance and low \ce{H}/\ce{O} ratio show a strong decrease in temperature from $2300$\,K to $1000$\,K between $1$ and $0.01$\,bar. \ce{P}+\ce{S} poor models with high \ce{H}/\ce{O} cool down to $1400-1600$\,K, then continue isothermally up to $10^{-5}$\,bar where they have a steep temperature inversion. All models rich in \ce{P}+\ce{S} have a temperature inversion between $10^{-2}$\,bar and $10^{-3}$\,bar. These atmospheres are strongly heated by \ce{CN} and \ce{PS}.
Figure \ref{fig:grid} shows that the most common condensates in \ce{N}-dominated atmospheres with \ce{C}/\ce{O}=2 are \ce{TiC(s,l)} and \ce{C(s)}. Likewise to the \ce{C}-dominated atmospheres we also find \ce{Al2O3(s,l)}, \ce{CaSiO3(s)}, \ce{Mg2SiO4(s,l)}, \ce{Fe(s,l)}, \ce{(CaAl)2SiO7(s)}, \ce{Ca2SiO4(s)}, \ce{MgO(s)}, and \ce{SiO(s)}. \ce{MgSiO3(s,l)} does not reach abundances $>1$\,ppb. Species which don't appear in \ce{C}-dominated atmospheres but are stable here are \ce{CaS(s)} and \ce{SiC(s)}, each appearing in $10\%$ of the cases and \ce{AlN(s)} condensing in $30\%$ of the models. The condensation is strongly affected by the sulphur and phosphorus content. In none of the \ce{P}+\ce{S} rich atmospheres (red and orange models) condensation can take place. In \ce{N}-dominated atmospheres with \ce{C}/\ce{O}=2, low \ce{P}+\ce{S} abundance and low \ce{H}/\ce{O} ratio (three darkest blue models), condensates form in the region of the temperature drop. In Figure \ref{fig:grid}, these species have lower occurrence in the \ce{N}-type atmospheres compared to the \ce{C}-atmospheres because only the three models with carbon and oxygen as major secondary volatiles in \ce{N}-atmospheres are cold enough, whereas all \ce{P}+\ce{S} poor models in the \ce{C}-dominated atmospheres are cold enough for condensation. In \ce{P}+\ce{S} poor models with high \ce{H}/\ce{O} ratio, \ce{SiC(s)} and \ce{AlN(s)} condense in the isothermal region between $0.1-10^{-5}$\,bar.

\subsubsection{\ce{H}-dominated atmospheres}
In \ce{H}-dominated atmospheres, there is a similar bifurcation in the temperature structures as in \ce{N}-dominated atmospheres. However, we find less stable condensates and we find them in fewer of the models compared to \ce{N}-dominated atmospheres at all \ce{C}/\ce{O} ratios. At low pressures, the non inverted atmospheres cannot cool down as efficiently as in the \ce{N}-dominated models because of increasingly reducing conditions favouring the formation of strong UV and optical absorbers. At low \ce{C}/\ce{O} ratios and high pressures, green house gas heating is more efficient in \ce{H}-dominated atmospheres, because more \ce{H2O} forms compared to the \ce{N}-dominated cases. At \ce{C}/\ce{O}=0.1 and \ce{C}/\ce{O}=0.5, a lot of oxygen is in \ce{O2} in \ce{N}-atmospheres. However in \ce{H}-atmospheres, \ce{H2O} is abundant and as a green house gas it heats up the near surface layers, resulting in a hotter deep atmosphere in these types.

In terms of condensates, we find stable species \ce{TiO2(s,l)}, \ce{SiO2(s,l)}, \ce{MgTi2O5(s,l)}, \ce{MgSiO3(s,l)}, \ce{MgAl2O4(s,l)}, \ce{Mg3(PO4)2(s,l)}, \ce{CaTiO3(s)} and \ce{Al2O3(s,l)}. Additionally, \ce{CaMgSi2O6(s)} and \ce{Mg2SiO4(s,l)} form if \ce{C}/\ce{O}=0.1 and \ce{Ti4O7(s,l)}, \ce{Ca2SiO4(s)}, \ce{(CaAl)2SiO7(s)}, and \ce{CaSiO3(s)} can be stable if \ce{C}/\ce{O}=0.5. The most common condensate in both sub-types is \ce{Al2O3(s,l)} forming in $50\%$ of the atmospheres with \ce{C}/\ce{O}=0.1 and in $40\%$ of the atmospheres with \ce{C}/\ce{O}=0.5.

\begin{figure*}
\centering
\includegraphics[width=0.98\textwidth]{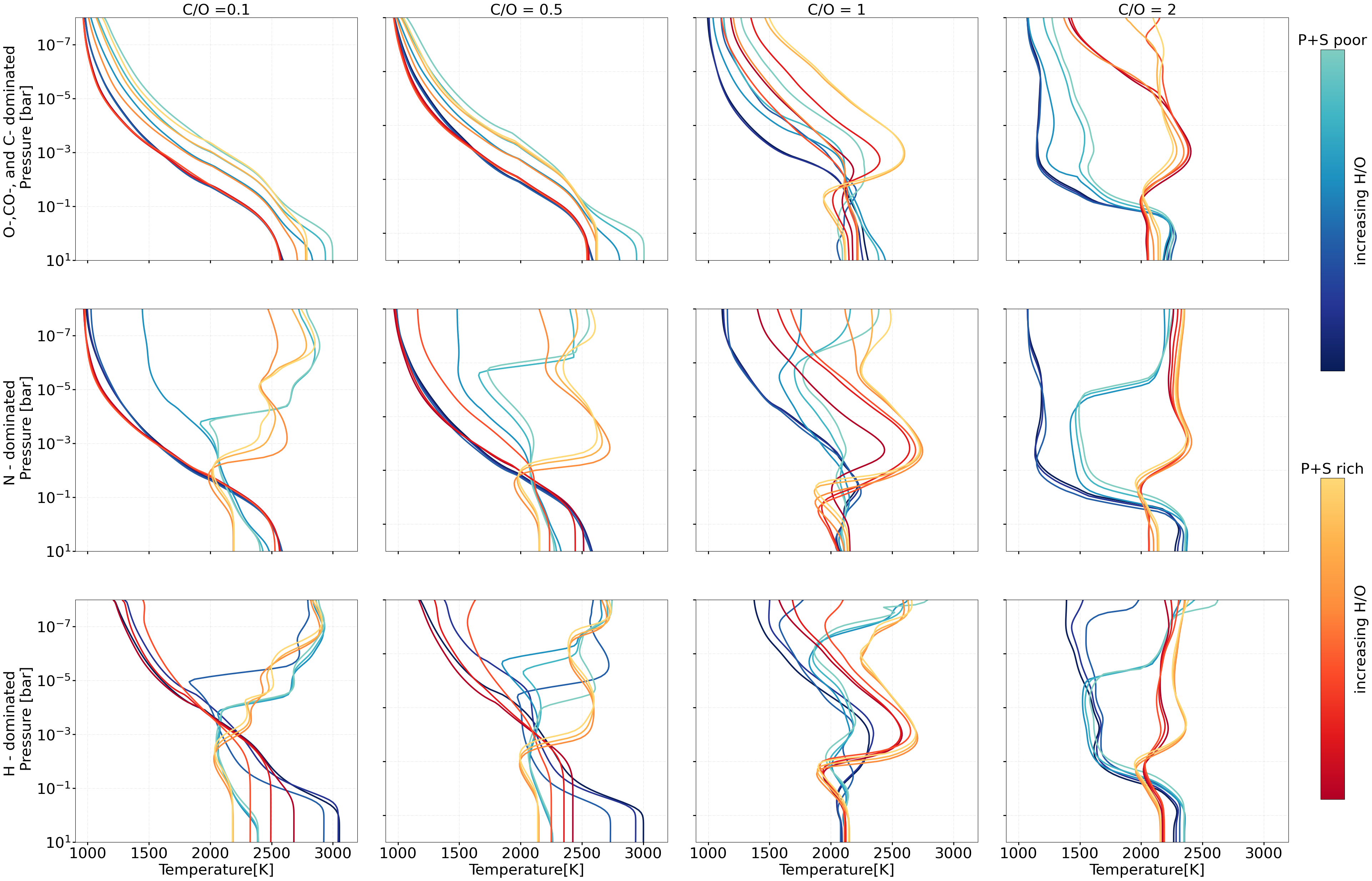} 
     \caption{Temperature structures of all models from our grid study. They are grouped according to their atmospheric type. The row indicates the dominating volatile in the atmosphere and the column indicates the \ce{C}/\ce{O} ratio of the model. The six blue and turquoise profiles correspond to models which have low \ce{P}+\ce{S} abundance. The six red and orange profiles have high \ce{P}+\ce{S} abundance. In each of these groups, the shade of the colour indicates the \ce{H}/\ce{O} ratio. The darker the blue/red shade, the lower the \ce{H}/\ce{O} ratio, the greener/yellower, the higher the \ce{H}/\ce{O} ratio. It is to note, that these colours do not indicate a numerical value of \ce{H}/\ce{O} ratio. The colour bar merely indicates a trend, but the lowest \ce{H}/\ce{O} ratio for \ce{H}-atmospheres is higher than the lowest \ce{H}/\ce{O} ratio for \ce{O}-atmospheres.}
     \label{fig:grid_TPs}
 \end{figure*}

In \ce{H}-dominated atmospheres with \ce{C}/\ce{O}=1, the temperature never drops below  $1400$\,K between $10-10^{-8}$\,bar, see Figure \ref{fig:grid_TPs}. Carbon and oxygen reside in the gas phase and no condensation can take place. It is to note, that this behaviour could potentially change in atmospheres with higher refractory abundances. 

\ce{H}-atmospheres with \ce{C}/\ce{O}=2, low \ce{P}+\ce{S} abundance and low \ce{H}/\ce{O} ratio show a strong decrease of temperature from $2300$\,K to $1600$\,K between $1-0.01 $\,bar. \ce{P}+\ce{S} poor models with high \ce{H}/\ce{O} (turquoise models) cool down to $1400$-$1600$\,K and are then isothermal up to $10^{-5}$\,bar before a steep temperature inversion. In the upper atmosphere, \ce{CN} creates the temperature inversion at  $5\cdot10^{-6}$\,bar. All the models with high \ce{P}+\ce{S} content have a temperature inversion between $10^{-2}$\,bar and $10^{-3}$\,bar due to the presence of \ce{PS}.
Overall, three condensates are stable in \ce{H}-atmospheres with \ce{C}/\ce{O}=2: \ce{C(s)} and \ce{TiC(s)} form in $50\%$ of the atmospheres and \ce{SiC(s)} forms in $25\%$ of the atmospheres. Likewise to \ce{N}-dominated atmospheres, the sulphur and phosphorus content affects the condensability of \ce{H}-dominated atmospheres with high \ce{C}/\ce{O} ratio. In \ce{P}+\ce{S} rich atmospheres nothing condenses. In atmospheres with low \ce{P}+\ce{S} abundance which are rich in carbon and oxygen, \ce{TiC(s)} can condense in the temperature drop between $1-0.01$\,bar. Graphite \ce{C(s)} is stable throughout most of the atmosphere. In \ce{P}+\ce{S} poor models which are also poor in carbon and oxygen, \ce{SiC(s)} is stable in the isothermal region in addition to \ce{TiC(s)} and \ce{C(s)}.

\subsubsection{Summary}
We have explored condensation in various atmospheric types which all have refractory abundances at solar values. Hence the conclusions we make are valid for these types only and might change with the inclusion of outgassing to our models.
Atmospheres which cool down the furthest at low pressures are the phosphorus and sulphur poor models with the lowest \ce{H}/\ce{O} ratio (dark blue models and \ce{O}-atmospheres). These are the cases in which we expect clouds to be the most stable, especially at high altitudes $P<0.01$\,bar. Depending on the atmospheric type, on the \ce{H}/\ce{O} ratio and on the \ce{C}/\ce{O} ratio, different solid and liquid species are stable. The \ce{H}/\ce{O} ratio is crucial in determining whether an inversion can occur. In atmospheres with \ce{C}/\ce{O}$<1$, \ce{O}-atmospheres are inversion free and the same holds for \ce{N}-atmospheres if they have a high oxygen content. For the compositions we investigate, the switch between an inversion and no inversions in \ce{N}-atmospheres with \ce{C}/\ce{O}$<1$ happens on the order of \ce{H}/\ce{O}=10 for \ce{P}+\ce{S} poor models and on the order of \ce{H}/\ce{O}=1 for \ce{P}+\ce{S} rich models. For \ce{C}/\ce{O}$\geq1$, the sulphur and phosphorus content are also important to consider. \ce{H}-, \ce{C}-, and \ce{N}-atmospheres with high sulphur and phosphorus content have inversions even if their \ce{H}/\ce{O} ratio is low. In terms of condensate stability at low \ce{C}/\ce{O} ratio, we predominantly find \ce{CaTiO3(s)} and \ce{TiO2(s,l)} in \ce{N}-atmospheres, \ce{CaTiO3(s)}, \ce{TiO2(s,l)} and \ce{SiO2(s,l)} in \ce{O}-atmospheres and \ce{CaTiO3(s)} and \ce{Al2O3(s,l)} in \ce{H}-atmospheres. Literature on the stability of condensates in atmospheres of Hot Jupiters (\ce{H}-atmospheres) by [e.g.] \citet{Wakeford2017} is in agreement with these findings. At high \ce{C}/\ce{O} ratio, liquid and solid oxides form only in specific cases where the bulk oxygen abundance is high enough and where the temperatures are low. These conditions can either be provoked by rainout of graphite or an initially large bulk oxygen abundance. Low sulphur and phosphorus abundances tend to result in lower temperature conditions compared to sulphur and phosphorus rich atmospheres because the heating agent \ce{PS} does not form in the former. Nitrides and carbides can be stable in the oxygen poorest atmospheres, i.e. where the \ce{H}/\ce{O} ratio and the \ce{C}/\ce{O} ratio are high.

\begin{figure*}
\centering
\includegraphics[width=0.90\textwidth]{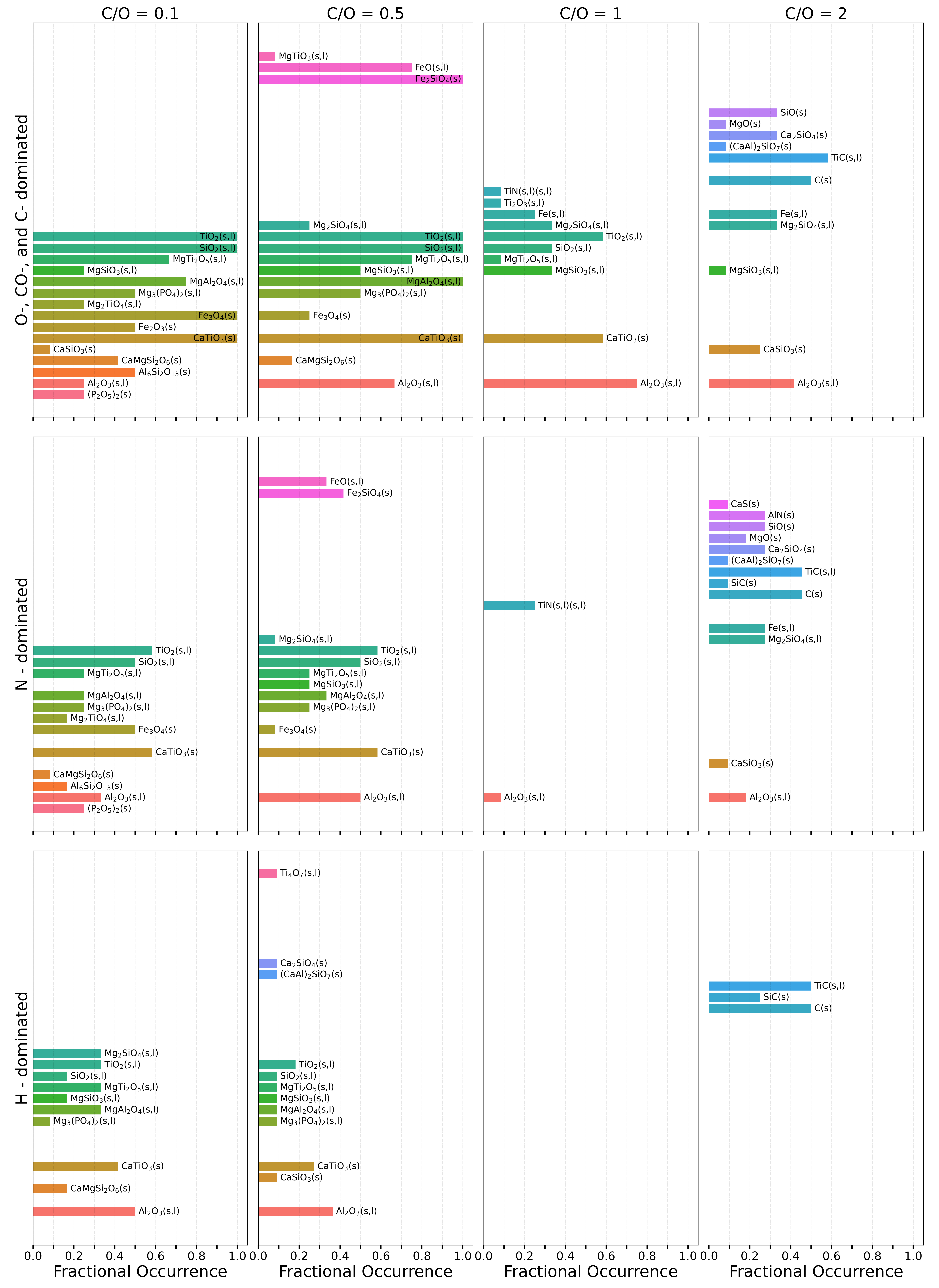}
     \caption{Shown are all condensates which appear in at least one model in the full grid and reach volume mixing ratios in the atmosphere $\geq1$\,ppb. Each panel represents one out of twelve atmospheric types: The row indicates the dominating volatile in the atmosphere and the column indicates the \ce{C}/\ce{O} ratio of the model. Each panel displays condensates as coloured bars, where each condensate has it's own attributed colour and the name is indicated in the bar. The x-axis shows the fraction of models in which each condensate appears for the corresponding atmospheric type.}
     \label{fig:grid}
 \end{figure*}

\subsection{Stability of condensates: The role of the oxygen budget}\label{obudget}

\begin{figure*}
    \centering
          \includegraphics[width=0.9\textwidth]{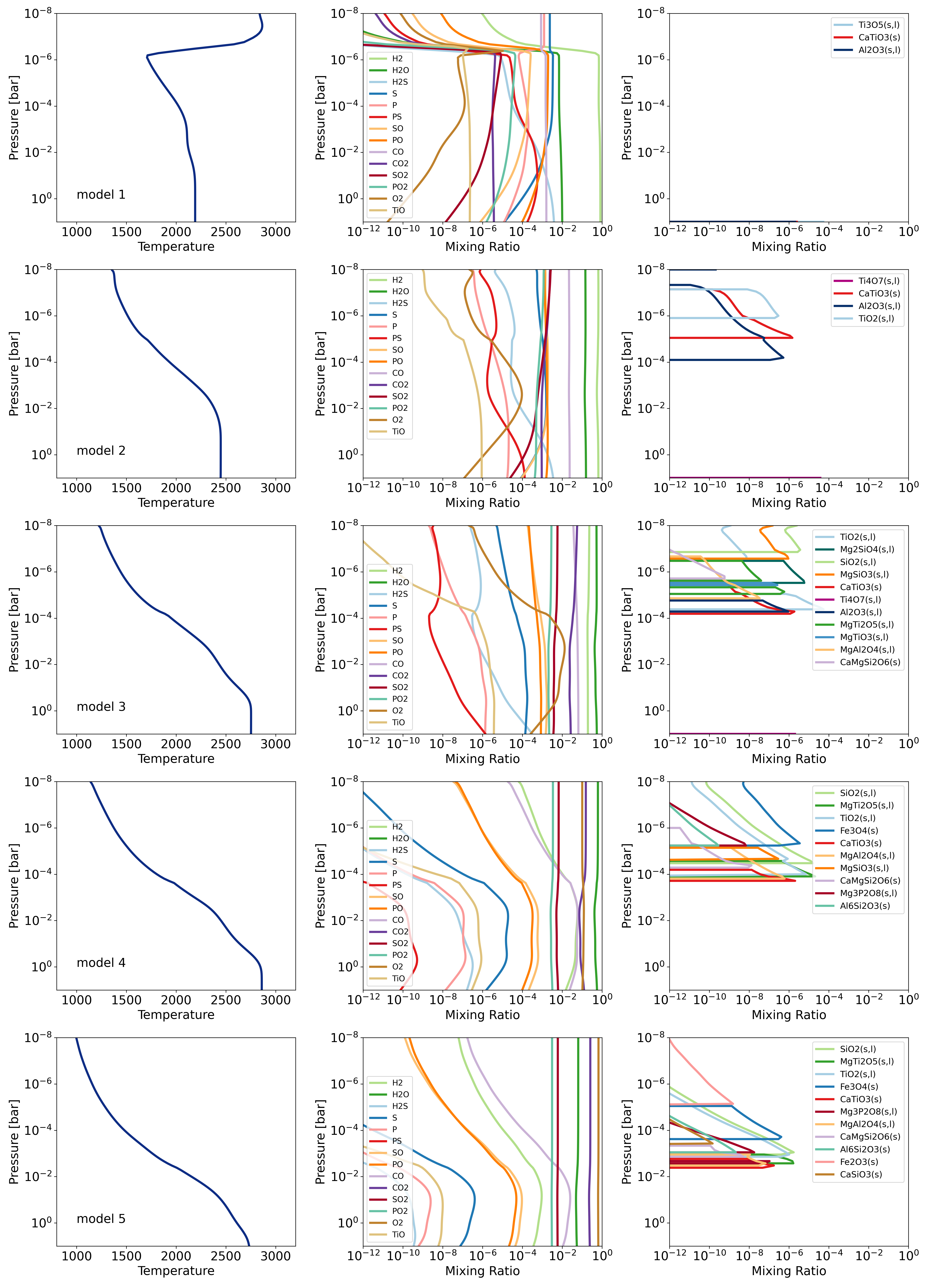}
\caption{Temperature profiles (left hand side column), relevant gas phase species (middle column) and condensates (right hand side column) of the five cases shown in Figure \ref{fig_piecharts}. The oxygen abundance is increased from top to bottom of the figure. In all cases, refractory elements are assumed to have solar abundances. The Figure highlights the cooling of the atmosphere and the increase in condensed species abundances with an increase in oxygen abundance.}
\label{fig_chemistry}
\end{figure*}

In this section, we explore in more detail the impact of the oxygen abundance of the atmosphere on condensate formation. The volatile abundances of the five compositions we consider for this study are the compositions described in section \ref{redox} and illustrated in form of pie charts in Figure \ref{fig_piecharts}. Figure \ref{fig_chemistry} shows the atmospheric gas-phase compositions, temperature–pressure profiles, and resulting condensates for this set of models. The oxygen abundance increases from left to right across the five models. The atmospheric composition transitions from a \ce{H2}-dominated atmosphere with \ce{O}/\ce{H}=0.0072 (model 1), to a \ce{H2}/\ce{H2O} mixture with \ce{O}/\ce{H}=0.011 (model 2), then to a \ce{H2O}-dominated atmosphere with \ce{O}/\ce{H}=0.46 (model 3), followed by a \ce{H2O}/\ce{CO2} atmosphere with \ce{O}/\ce{H}=0.93 (model 4), and finally to an \ce{O2}/\ce{CO2}-dominated atmosphere with \ce{O}/\ce{H}=15 (model 5). The amount of available oxygen affects the type of condensates that can form in two ways:\\
Thermal Structure: The \ce{O}/\ce{H} ratio and the \ce{C}/\ce{O} ratio have a strong influence on the thermal structure of the atmosphere (see section \ref{condensates}): At extremely high \ce{O}/\ce{H} ratios and low \ce{C}/\ce{O} ratio - dark blue and red models in the two first columns of Figure \ref{fig:grid_TPs}- , we expect high surface temperatures and then a decrease of temperature with altitude which reaches temperatures $<1300$\,K between $10^{-5}-10^{-8}$\,bar. This favours cloud formation in those regions. At a lower \ce{H}/\ce{O} ratio, major optical absorbers (e.g. \ce{TiO}, \ce{CN} and \ce{PS}) can form depending on the composition and heat up the atmosphere, creating inversions and preventing cloud formation. \\
Condensate Formation: Higher oxygen abundances promote the formation of a greater number of condensates. As shown in Figure \ref{fig:grid}, the most common condensates are oxides. In chemical equilibrium, the formation of condensates is governed by the thermal stability of all gaseous, liquid, and solid species and the species which are the most stable will condense first. Nothing condenses in the atmosphere of model $1$, which is dominated by molecular hydrogen see Figure \ref{fig_chemistry}. However, part of the atmosphere rains out to the surface in the form of \ce{CaTiO3(s)}, \ce{Al2O3(s)} and \ce{Ti3O5(s)}. The inversion is caused by strong short-wave absorbers, mainly \ce{PS}, which can form in this extremely reducing environment. This is similar to the \ce{P}+\ce{S} rich cases of \ce{H}-dominated atmospheres discussed in section \ref{condensates}.
In an atmosphere with a slightly higher oxygen budget - model $2$, \ce{H2} is still the major gas phase component. Phosphorus and sulphur form \ce{PO}, \ce{SO}, \ce{SO2} and \ce{PO2}, \ce{S} and \ce{P} instead of \ce{PO}, \ce{SO}, \ce{S}, \ce{P}, \ce{HS} and \ce{PS} like in model $1$. Hence, the temperature inversion has disappeared from one model to the other. This allows the most thermally stable condensates to form. As shown in Figure \ref{fig_chemistry}, these are in order at which they form from bottom to top of the atmosphere: \ce{Al2O3(s)}, \ce{CaTiO3(s)} and  \ce{TiO2(s,l)}. \ce{Ti4O7(s)} is also thermally stable under the given conditions, but it rains out at the planetary surface. These particularly stable species are also the ones which can form in most of the compositions we explore in our grid in section \ref{condensates}.
As the oxygen budget of the atmosphere is increased to \ce{O}/\ce{H}=0.46 in model $3$, \ce{CaTiO3(s)} and \ce{Al2O3(s)} remain the dominating candidates forming at the highest temperatures. At slightly colder temperatures, Mg-oxides such as \ce{Mg2SiO4(s,l)} and \ce{MgTi2O5(s,l)} form, followed by \ce{SiO2(s,l)}, \ce{MgSiO3(s,l)} and \ce{TiO2(s,l)}. Iron only condenses in oxidised form when the temperatures are low and the oxygen budget is high. A fraction of \ce{O}/\ce{H}=0.46 in models $3$ is not enough for stable iron oxides, but with \ce{O}/\ce{H}=0.93 in model 4, \ce{Fe3O4(s,l)} becomes stable at pressures $<10^{-5}$\,bar. In model $5$, \ce{Fe3O4(s,l)} is stable between pressures of $10^{-3.5}$ to $>10^{-5}$\,bar, where it is replaced by its more oxidised version \ce{Fe2O3(s,l)}. 
This behaviour can be traced back to the dependence of the stability of solid and liquid oxides on oxygen abundance as well as on pressure and temperature conditions. \citet{Seidler2025} investigate this behaviour for a range of condensates. They find that silicates and magnesium- oxides can condense first at mediate oxygen fugacities. Similarly to what our models predict, they also conclude that the iron condensates \ce{Fe3O4(s)} and the more oxidised form \ce{Fe2O3(s)} only form at particularly high oxygen fugacities.

One factor which can raise the condensation temperature is enhancement of the abundance of the related vapour species to form certain condensates. Similarly to  how an increased oxygen abundance can favour the condensation of oxides. If the condensing species' abundance is raised, the condensation curve can move towards higher temperatures, and vice versa if its abundance decreases. This means, that in general, the condensation temperature of refractory rich clouds is higher for metal rich atmospheres than for metal poor cases.

\subsection{A more complex cloud model reveals thicker clouds} \label{exolyn}

To explore how different cloud formation models affect the results, we compare the composition and vertical extent of clouds using the dynamical cloud model {\textsc{ExoLyn}} and contrast these with our previous chemical equilibrium calculations using rainout. We use the \ce{O2}/\ce{CO2} atmosphere (model $5$) for this comparison, as it produces the largest amount of condensates. We also explore a carbon rich scenario using model $5$ as a baseline and enhancing the \ce{C}/\ce{O} ratio. To highlight the differences of two extremes we choose a high eddy diffusion coefficient of K$_{\text{zz}}=10^{10}$\,cm$^2$s$^{-1}$ in the cloud formation model for our comparison. We then investigate in greater depth the effect of vertical mixing on cloud extent and particle size in a separate section. We include cloud species and their condensation reactions as listed in Table \ref{Tab_reactions}.

\subsubsection{Clouds in \ce{O}-dominated atmospheres}\label{O2_comparison}

\begin{figure*}
        \centering
        \begin{subfigure}[b]{0.98\textwidth}
        \centering
           \includegraphics[width=0.49\textwidth]{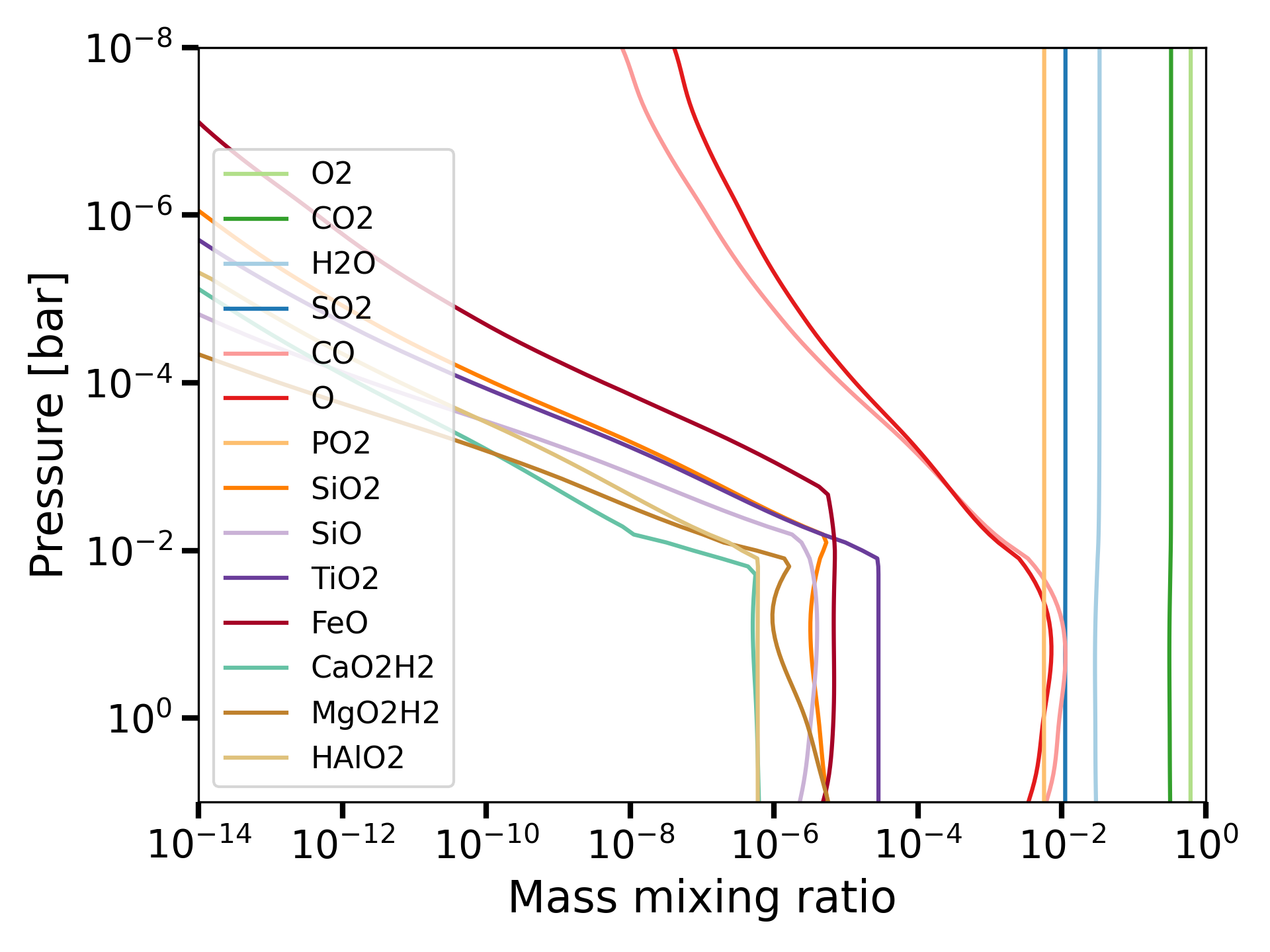}
         \includegraphics[width=0.49\textwidth]{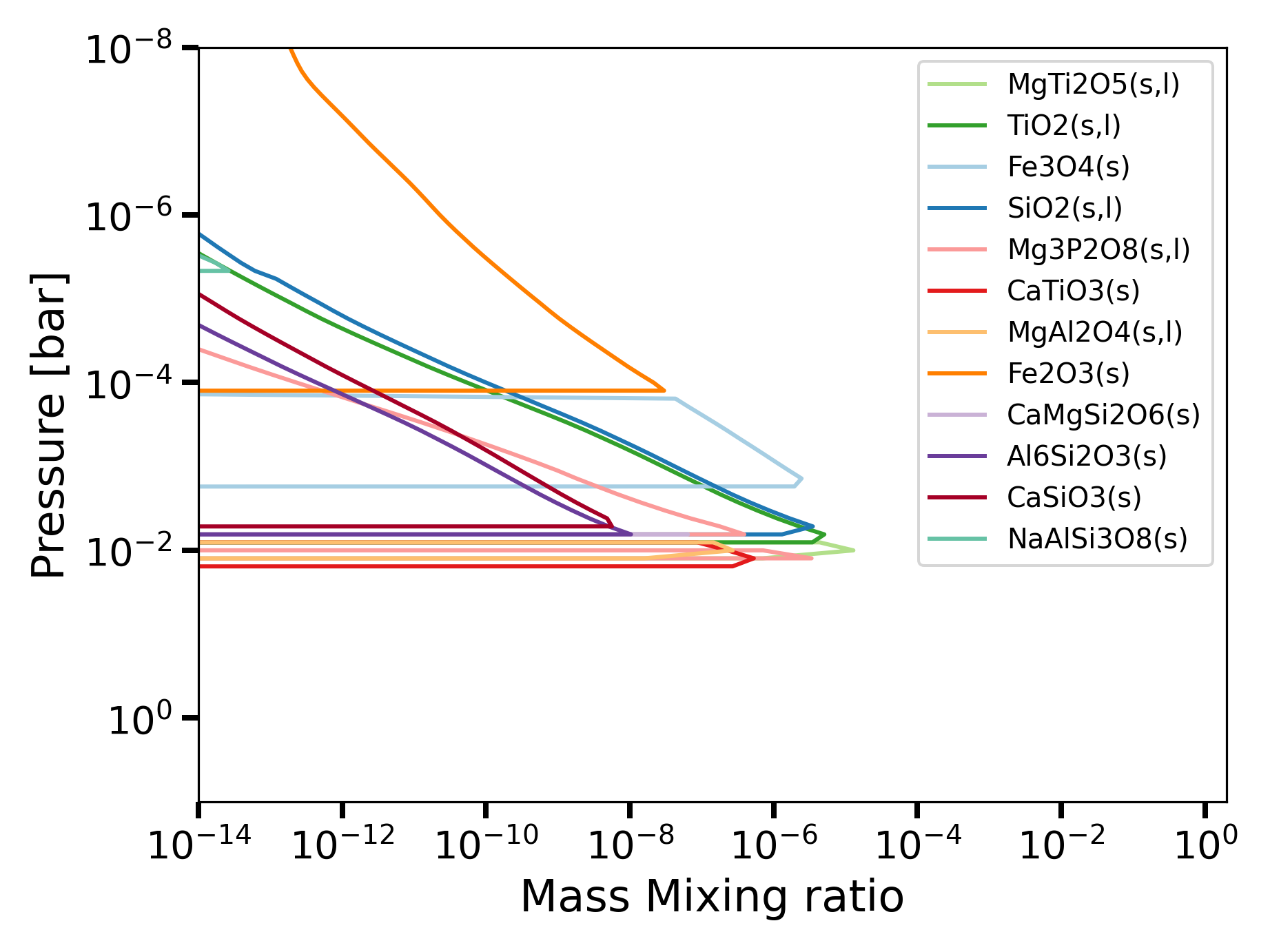}
    \end{subfigure}
    \begin{subfigure}[b]{0.98\textwidth}
        \centering
        \includegraphics[width=0.49\textwidth]{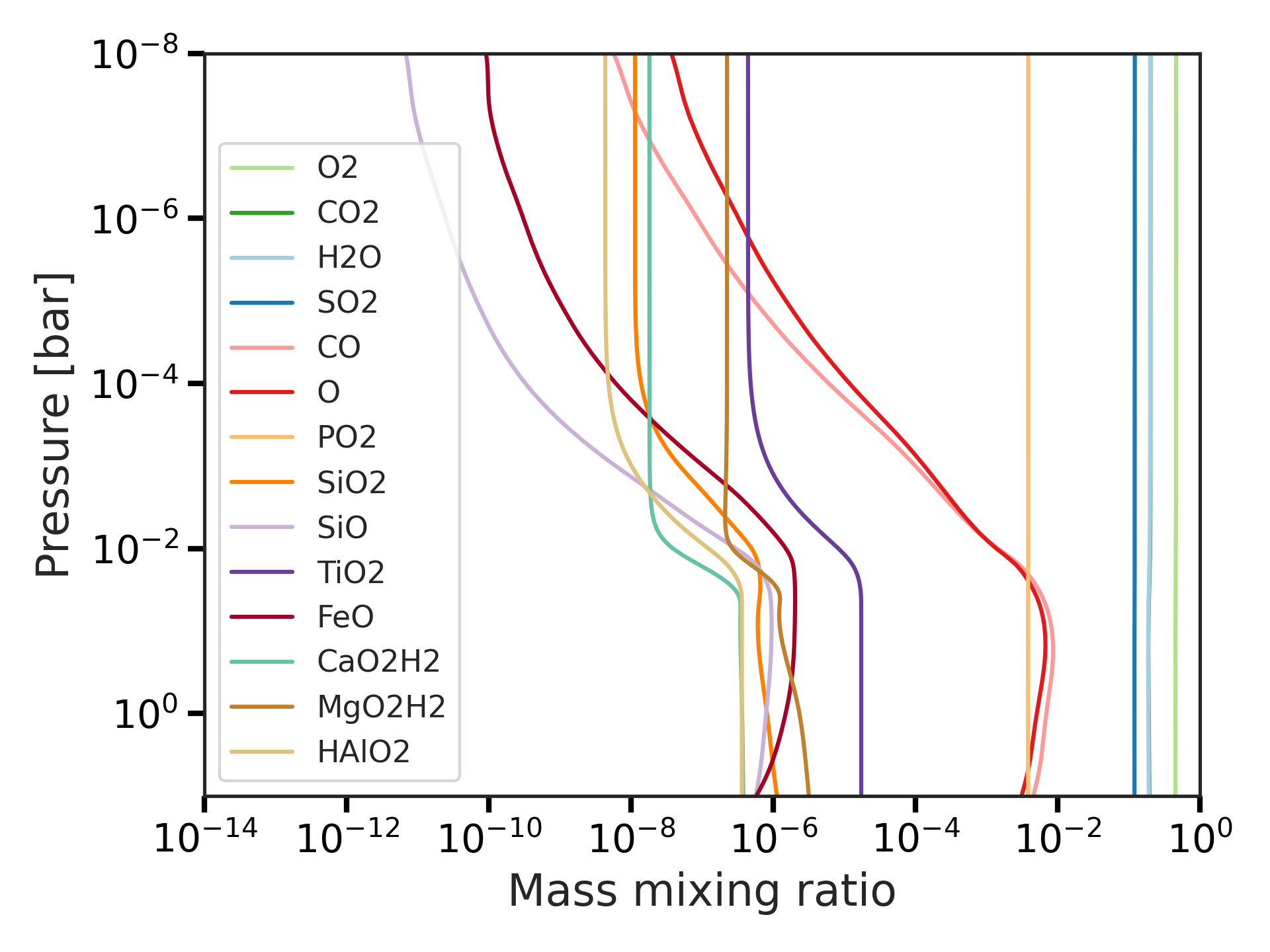}
        \includegraphics[width=0.49\textwidth]{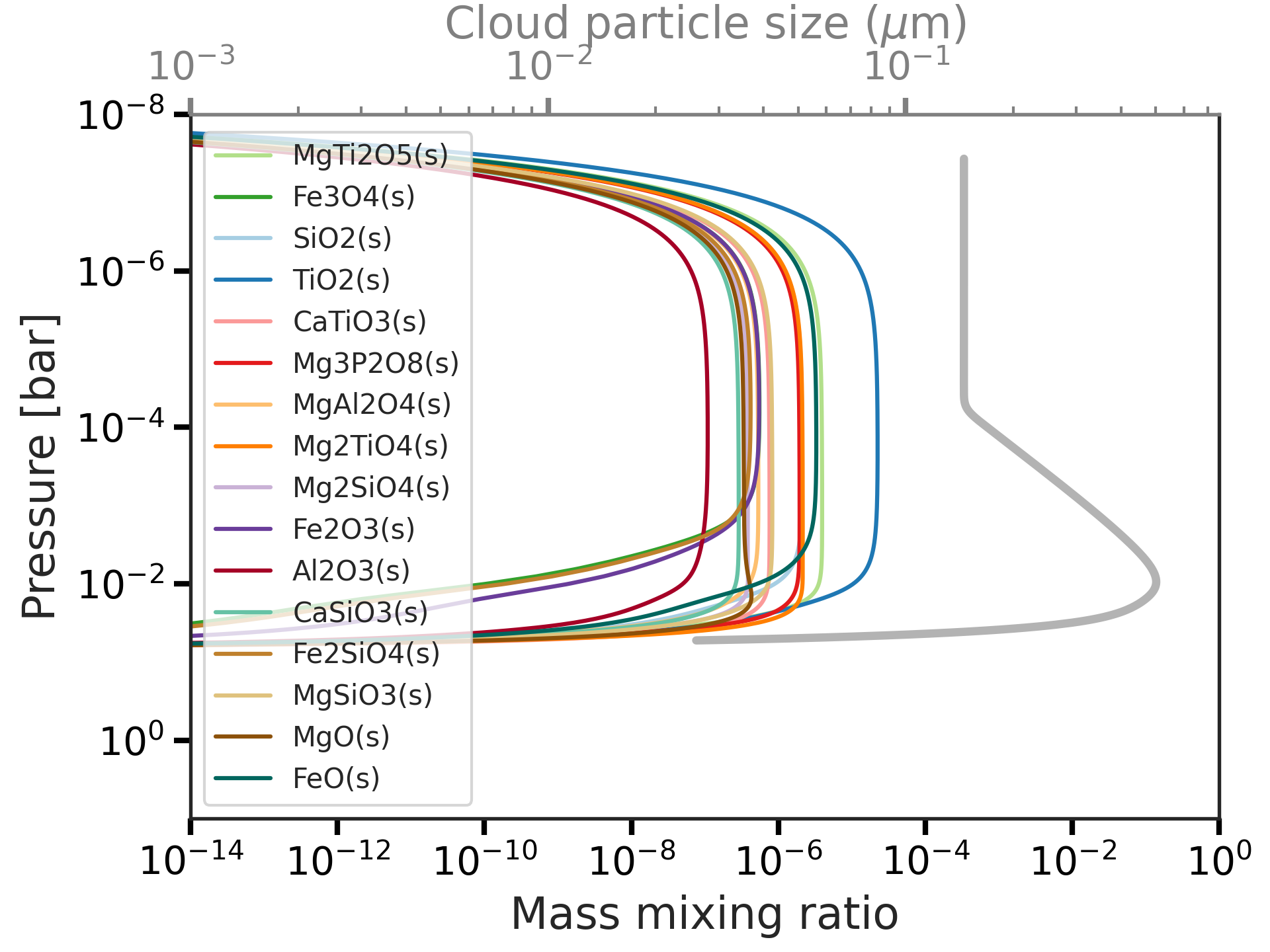}
    \end{subfigure}
        \caption{Mass mixing ratios and condensate/cloud of the \ce{O2}/\ce{CO2} atmosphere - model 5. The $15$ major gas phase species are shown in the left hand side column and the solid and liquid species in the right hand side column. The upper row shows the chemistry computed with the rainout approach with {\textsc{FastChem}}3.For comparative purposes to {\textsc{ExoLyn}} we have converted the standardised output of volume mixing ratios to mass mixing ratios. In the bottom row, we have computed the cloud mass mixing ratios, particle sizes and the gas phase with the cloud formation model {\textsc{ExoLyn}} and {\textsc{FastChem}}3 combined as described in section \ref{exolyn}. The grey curve in the cloud panel indicates the particle size, and the coloured lines represent the mass mixing ratios of the individual cloud components.}
\label{fig_Crich_oxi5}
\end{figure*}

Figure \ref{fig_Crich_oxi5} shows how the cloud in model $5$ predicted with {\textsc{ExoLyn}} compares to the rainout scenario. With both approaches, maximal cloud masses $>10$\,ppm are reached. One difference between the two cases is the extent of the cloud. The mass of the cloud in the rainout model is concentrated at the bottom layers. It decreases drastically with altitude above $10^{-4}$\,bar. The bottom cloud layer, below which material starts to evaporate is situated around $0.01$\,bar. The cloud in the dynamical model reaches mass mixing ratios higher than $10$\,ppm between $10^{-2}-10^{-6.5}$\,bar and stretches to 10$^{-7.5}$\,bar. The bottom of the cloud is located at pressures $>0.01$\,bar. The most stable solid species in the rainout model, evaporating at the hottest temperatures is \ce{CaTiO3(s)}. \ce{MgTi2O5(s,l)} is stable around $10^{-2}$\,bar with a mass mixing ratio of $5- 10$\,ppm depending on the altitude. It is replaced by $0.1$\, ppm \ce{Mg3(PO4)2(s,l)} and $5$\,ppm \ce{TiO2(s,l)} and \ce{SiO2(s,l)} above $10^{-2}$\,bar. In the dynamical model, \ce{Mg2TiO4(s)} is most abundant at the cloud bottom, where it reaches mass mixing ratios $>1$\,ppm. At $\approx0.05$\,bar, \ce{TiO2(s)} condensation becomes stable making the latter the major cloud forming material with \ce{MgTi2O5(s)} as the second most abundant species. Two factors enhance the \ce{MgTi2O5(s)} extent in the dynamical model compared to the equilibrium model:

Mass conservation in the rainout approach: In the rainout approach, the atmospheric mass is not conserved in the layers above the condensate because all condensed elements are removed from the atmosphere when they rain out. In this case, \ce{MgTi2O5(s)} raining out results in a change of composition above each layer where \ce{MgTi2O5(s)} is stable. With decreasing pressure, the temperature also decreases and additional species such as \ce{TiO2(s,l)} and \ce{Mg3(PO4)2(s,l)} become stable, replacing \ce{MgTi2O5(s)}. In reality, the material between the different atmospheric layers would be mixed, and all of these species would condense at the same time.

Non-equilibrium processes in the dynamical model: A combination of the different dynamical cloud formation processes leads to abundances deviating from chemical equilibrium. In {\textsc{ExoLyn}}, sedimentation and vertical mixing are competitive processes affecting the cloud particle size and the cloud extent. Particles which form at lower altitudes can be transported to higher atmospheric layers. Therefore, \ce{MgTi2O5(s)} can stay stable throughout the upper layers of the atmosphere, if the conversion from \ce{MgTi2O5(s)} to \ce{TiO2(s)} happens very slowly. In addition to this time dependence of the mechanisms, nucleation can also promote formation of solid and liquid species in regimes where equilibrium alone does not allow condensation. The reason for this is that the presence of condensation nuclei provides a surface onto which material can condense and accumulate in solid or liquid form. Additional magnesium bearing species can form in the dynamical model. Almost $1$\,ppm of the cloud mass is locked into \ce{MgSiO3(s)} and \ce{Mg2SiO4(s)} even though both of them are not abundant in chemical equilibrium. \ce{TiO2(s)} is the dominating cloud component in the dynamical cloud model throughout most of the cloud, with mass mixing ratios $\approx50$\,ppm. At pressures $\leq10^{-2.5}$\,bar, \ce{FeO(s)} and \ce{MgTi2O5(s)} are the second most abundant cloud components after \ce{TiO2(s)} and reach mass mixing ratios of $\approx5$\,ppm. In the rainout model, iron condenses mainly as \ce{Fe2O3(s)} and \ce{Fe3O4(s)} at pressures smaller than $5\cdot 10^{-2}$\,bar. In the dynamical model, these two species are abundant in this pressure range as well with mass mixing ratios of $\approx0.5-1$\,ppm, where we also find \ce{Fe2SiO4(s)} with an abundance of $\approx0.5$\,ppm. In both models, \ce{SiO2(s)} is the major silicon reservoir of the cloud grains.

The comparison between the two models shows that the choice of the model matters. The mass of the predicted cloud, its extent, and its composition depend on this choice. Because material is taken out of the atmosphere in every layer with rainout, the available material for condensation is limited and decreases with increasing altitude. As a result, the rainout model underestimates the extent and the mass of the cloud. In addition, the predicted cloud composition can also be inaccurate because the species which are only thermodynamically stable deep in the atmosphere cannot be transported upwards. The dynamical cloud model can present a more accurate estimate of the cloud mass, extent, and composition but requires knowledge of the nucleation rate and mixing strength in the planetary atmosphere.
Equilibrium chemistry alone does not provide cloud particle sizes, however the cloud model does. These are important to predict the extent of the cloud, since larger particles settle more efficiently than small particles. In our cloud model with vertical transport of K$_{\text{zz}}=10^{10}$\,cm$^2$s$^{-1}$, the particle sizes range between $0.1-0.2$\,$\mu$\,m, with larger particles at greater depths because setteling is the strongest there. Different particle sizes also lead to different spectral features, since the particle size determines which wavelengths get scattered the strongest. We will discuss spectral features in section \ref{kcoefficients}. One source of uncertainty for the grain composition in the dynamical cloud model which could also explain some of the discrepancies between the two approaches is the fact that the distribution of elements in the mixed cloud is influenced by the reaction pathways we take into account with {\textsc{ExoLyn}}.

\subsubsection{Clouds in \ce{C}-dominated atmospheres}

\begin{figure*}
        \centering
        \includegraphics[width=0.33\textwidth]{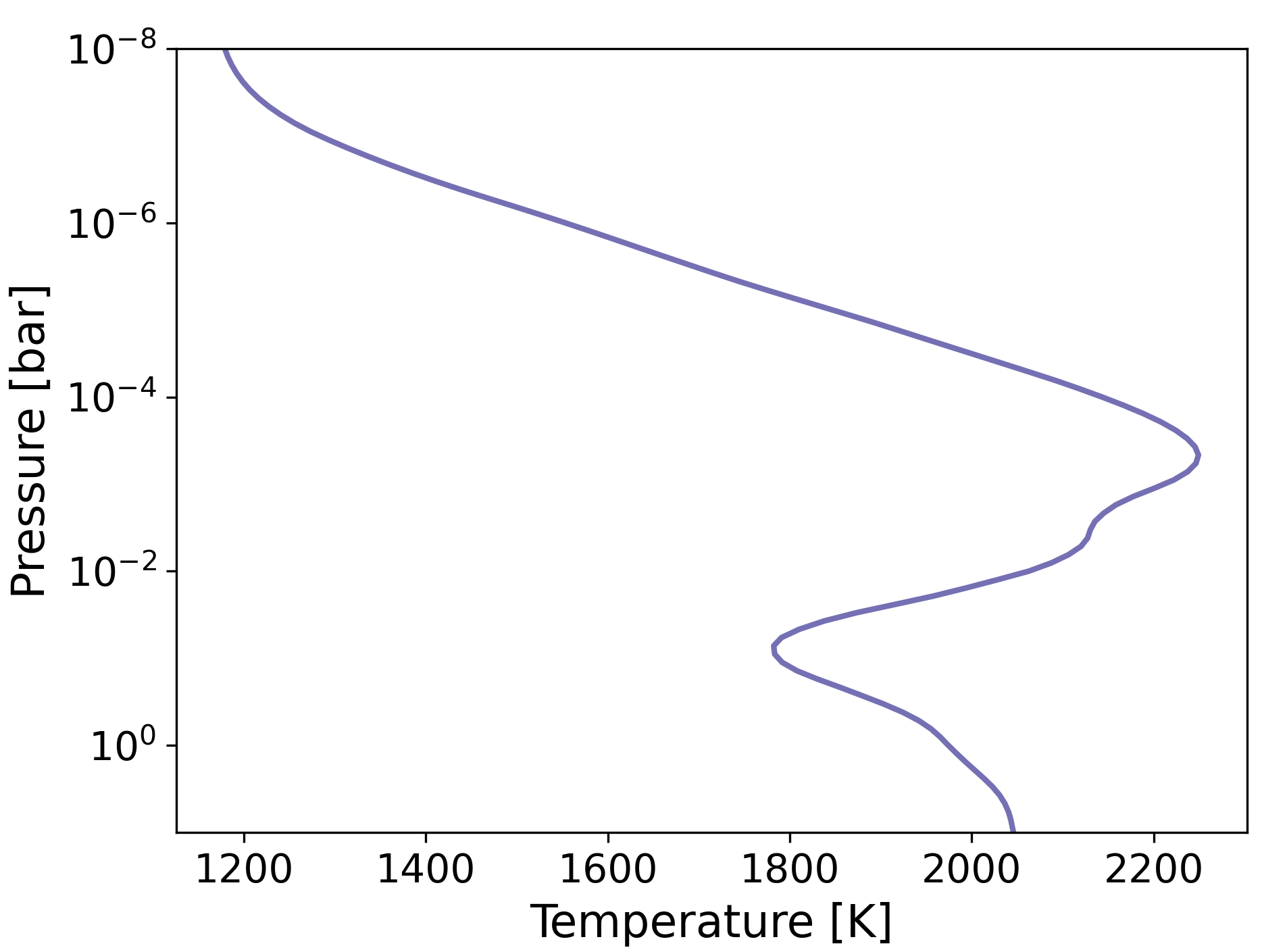}
        \includegraphics[width=0.33\textwidth]{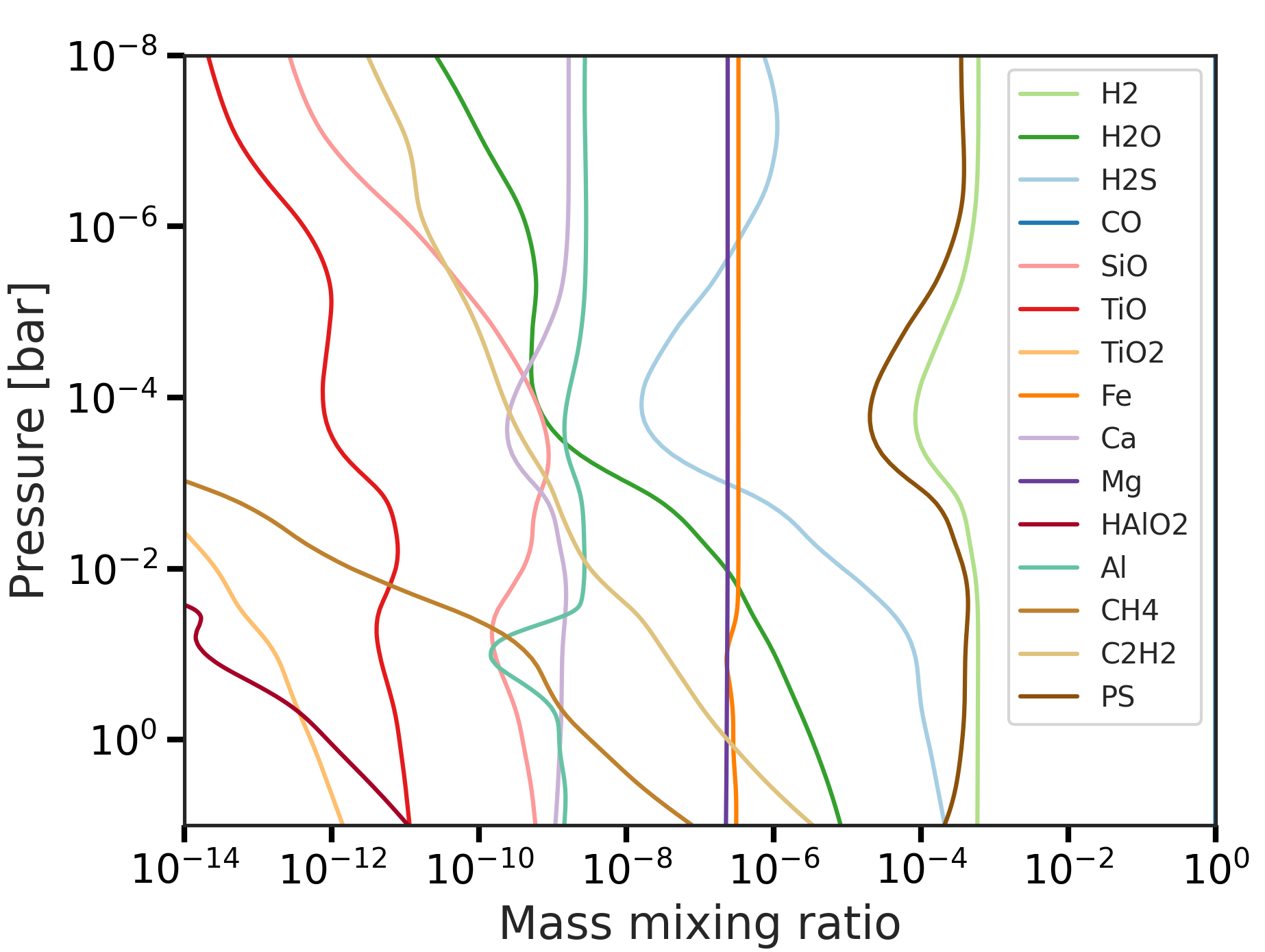}
        \includegraphics[width=0.33\textwidth]{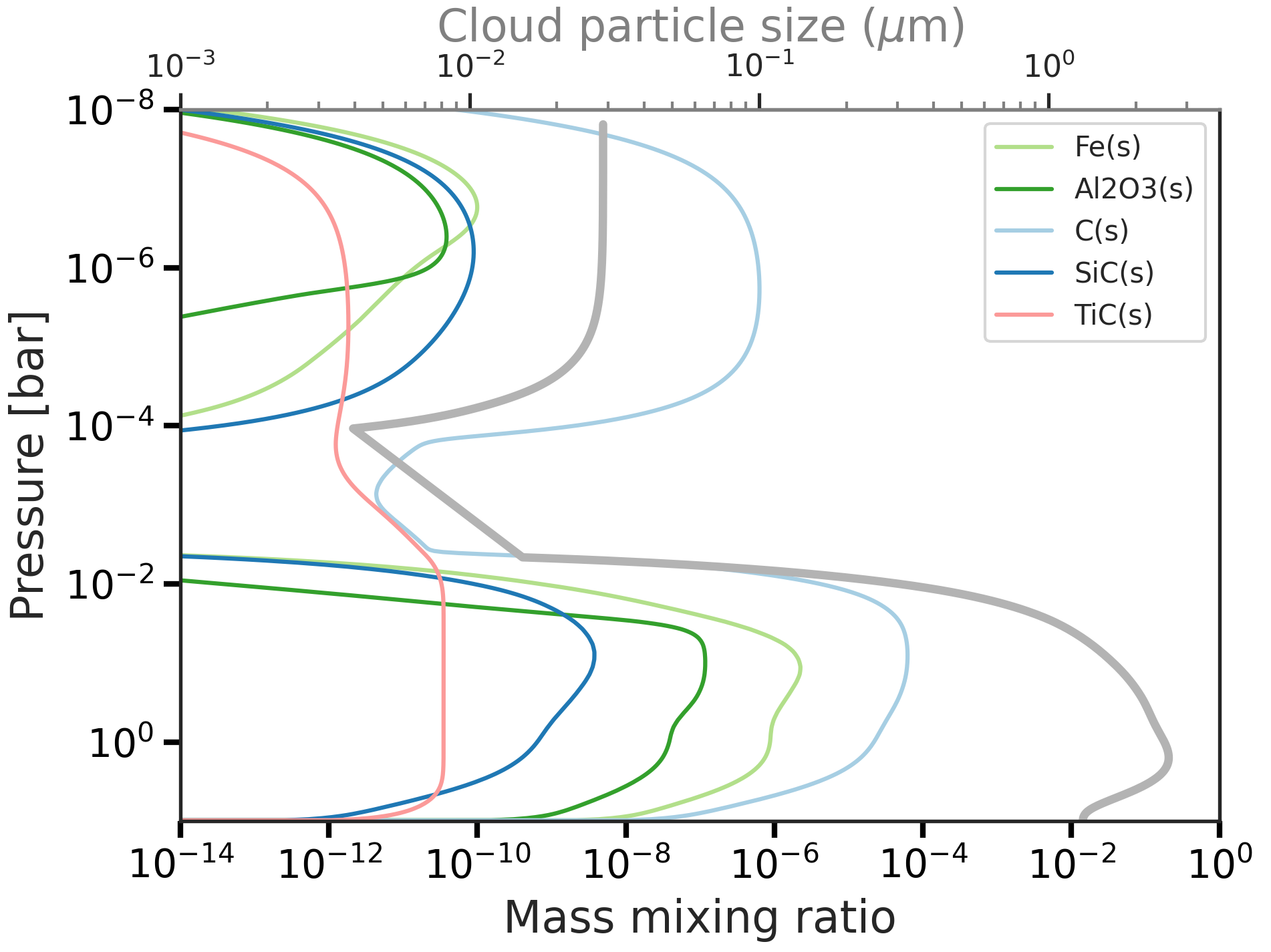}
        \caption{\ce{C}-atmosphere which is like model 5 but with \ce{C}/\ce{O}=2. Shown are temperature structures in the first column, $15$ of the important  mass mixing ratios of the gas phase in the second column and mass mixing ratios of the cloud components in the third column. We have computed the cloud mass mixing ratios, particle sizes and the gas phase with the cloud formation model {\textsc{ExoLyn}} and {\textsc{FastChem}}3 combined as described in section \ref{exolyn}. The grey curve in the cloud panel indicates the particle size, and the coloured lines represent the mass mixing ratios of the individual cloud components. The reaction rates we used are listed in Table \ref{Tab_reactions}.}
\label{fig_Crich}
\end{figure*}

The reaction pathways for clouds in carbon rich environments are poorly studied. We select the reaction pathways for \ce{TiC(s)}, \ce{SiC(s)} and \ce{C(s)} from proposed reactions by \citet{Helling2017} according to which pathways result in the largest cloud mass. In sections \ref{condensates} and \ref{obudget} we have demonstrated that the cloud composition and extent are both affected by the \ce{C}/\ce{O} ratio, the abundance of volatiles and the temperature and pressure. In this section we investigate how a higher \ce{C}/\ce{O} ratio could affect the atmosphere and the clouds. We construct for this purpose a \ce{C}-atmosphere by taking the example case of Figure \ref{fig_piecharts} model 5 discussed in section \ref{O2_comparison}. We keep the \ce{HNPS} abundances but set the \ce{C}/\ce{O} ratio to \ce{C}/\ce{O}=2. Figure \ref{fig_Crich} shows temperature structure, gas composition and cloud composition of this atmosphere.

The gas composition is drastically different from the \ce{O2}/\ce{CO2}-atmosphere. The major gaseous component is \ce{CO} reaching almost a mixing ratio of unity. Part of the graphite has rained out to the surface. Nevertheless a thick graphite cloud still extends from the near surface region to the upper most layers of the atmosphere. Mixed into the graphite grains are \ce{Al2O3(s)}, \ce{Fe(s)} and \ce{TiC(s)}. The cloud forms a two layer structure, because of the shape of the temperature profile. The grain sizes are particularly large in the lower layer in the near surface regions, reaching diameters $>1$\,$\mu$m between $10-0.01$\,bar. This lower layer ends above $0.01$\,bar with the first temperature inversion. A second layer reappears around $10^{-4}$\,bar, where the atmosphere has cooled down again. 
Overall, the thick carbon cloud deck dominated by graphite agrees with our finding in the grid study section \ref{condensates}, that graphite clouds are a particularly stable cloud species in \ce{C}-dominated atmospheres.

\subsection{Spectral features of highly refractory clouds} \label{kcoefficients}

\subsubsection{Optical properties of dust grains}

 \begin{figure*}
    \centering
        \includegraphics[width=0.99\textwidth]{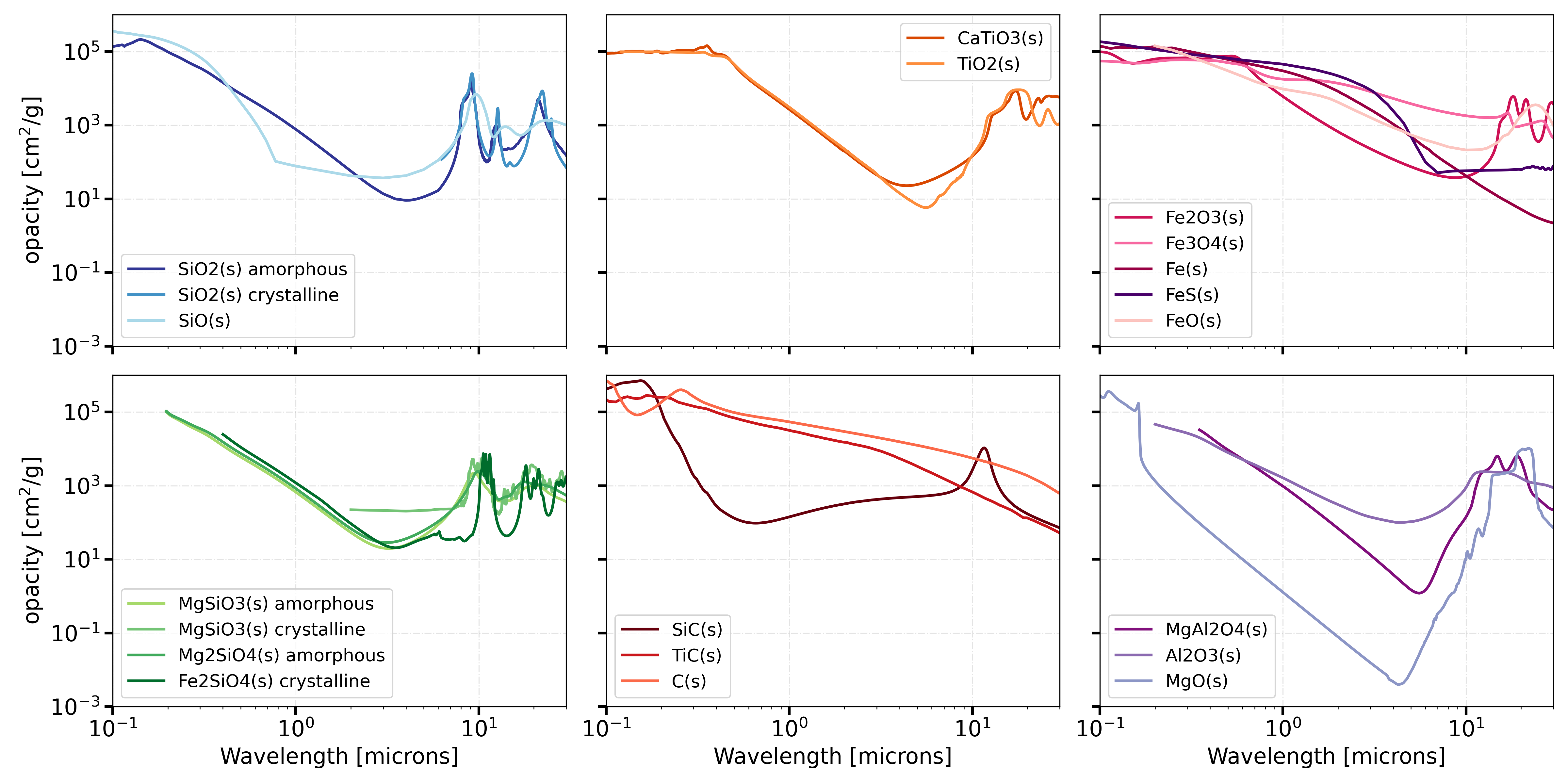}\\
    \caption{Extinction coefficient of solids as calculated with {\textsc{optool}} for a $0.01$\,$\mu$m large particle. We show all species in our models for which we could find refractive index data. The species are grouped according to their spectral features. For the species with a graph not spanning the full wavelength range between $0.1-30$\,$\mu$m, part of the optical and UV data is missing. All the sources we use for the optical data here are listed in Table \ref{opconst}.}
    \label{nk}
\end{figure*}

In this section we identify the potential features of all dust particles which appear in our grid Figure \ref{fig:grid} and for which optical data is available. We use these features to analyse our spectra in section \ref{spectra_oxygen} and  section \ref{spectra_CH} and to provide a comparison between spectral features of individual dust species and mixed cloud grains. In Figure \ref{nk} we show the extinction coefficients for the most common condensates in our model, computed as described in section \ref{grid_methods}. We show both amorphous and crystalline compounds if both are available (e.g. for \ce{SiO2(s)} and \ce{MgSiO3(s)}), because the structure of solid or liquid which the different materials can take is highly uncertain \citep[e.g.][]{Moran2024}. \citet{Wakeford2015} show extinction coefficients for some of the cloud species considered in this paper and conclude that the individual species will be difficult to discern, but they can be grouped according to their features and these groups can be identified. We also group the extinction coefficients according to the vibrational bands of the solids. As shown in Figure \ref{nk}, some optical data is missing. There is a lack in refractive index data at wavelengths shorter than $0.4 $\,$\mu$m for crystalline \ce{Fe2SiO4(s)} and for \ce{MgAl2O4(s)}. For crystalline \ce{SiO2(s)} data is missing up to $5$\,$\mu$m, for crystalline \ce{MgSiO3(s)} up to $1.1 $\,$\mu$m and for amorphous \ce{MgSiO3(s)} and \ce{Mg2SiO4(s)} as well as for \ce{Al2O3(s)}, refractive indices are available for wavelengths larger than $0.2 $\,$\mu$m only.

The features of the small grains can be split into two categories: The UV and optical features (slope) and the features in the infrared. The species which could be identifiable from infrared features are \ce{SiO2(s)}, \ce{SiO(s)}, \ce{Fe2SiO4(s)}, \ce{CaTiO3(s)}, \ce{TiO2(s)}, \ce{Al2O3(s)}, \ce{MgAl2O4(s)}, \ce{MgO(s)}, \ce{Fe2O3(s)}, \ce{Fe3O4(s)}, \ce{MgSiO3(s)}, \ce{Mg2SiO4(s)} and \ce{MgO(s)}. \ce{SiC(s)}, \ce{C(s)} and \ce{TiC(s)}, \ce{Fe(s)}, \ce{FeS(s)} and \ce{FeO(s)} have a slightly less characteristic behaviour in the infrared part of their extinction coefficients. Distinct features in the optical and UV can arise from \ce{TiO2(s)}, \ce{CaTiO3(s)}, \ce{SiO2(s)}, \ce{Fe2O3(s)}, \ce{MgO(s)}, \ce{SiC(s)} and potentially other species for which we are missing UV data. The other materials for which we have data would mainly have a flattening effect on the spectrum and change the slope in the UV and optical compared to a purely gaseous atmosphere.

\subsubsection{Spectral imprints of clouds in an \ce{O2}/\ce{CO2} atmosphere}\label{spectra_oxygen}

As discussed in section \ref{condensates}, \ce{O}-dominated atmospheres are unlikely to have temperature inversions and are therefore favourable environments for high altitude cloud formation. In the following, we investigate whether clouds could be observable in \ce{O}-dominated atmospheres of small planets. For all silicates we have used the amorphous properties to compute cloud opacities, since these are the most complete ones in wavelength space. However, we note that the features from crystalline structures could potentially be more distinct than those of their amorphous counterparts. In addition, experiments suggest a crystalline rather than an amorphous structure at high temperatures $T>800$\,K \citep{Zeidler2013,Henning1997,Kitzmann&Heng2018}. For our purposes, it is still a good enough approximation to take the amorphous properties due to the lack in crystalline data and due to the multiple structures which the crystal lattices can take depending on the exact atmospheric conditions (temperature, pressure and composition) \citep{Moran2024}. In addition, \citet{Moran2024} highlight, that is also unknown under which exact condition, liquid \ce{SiO2(l)} drops could form, resulting in "glassy", amorphous like features.  Investigating the effects of different lattice structures on observability is beyond the scope of this paper.

\begin{figure*}
    \centering
        \centering
        \includegraphics[width=0.98\textwidth]{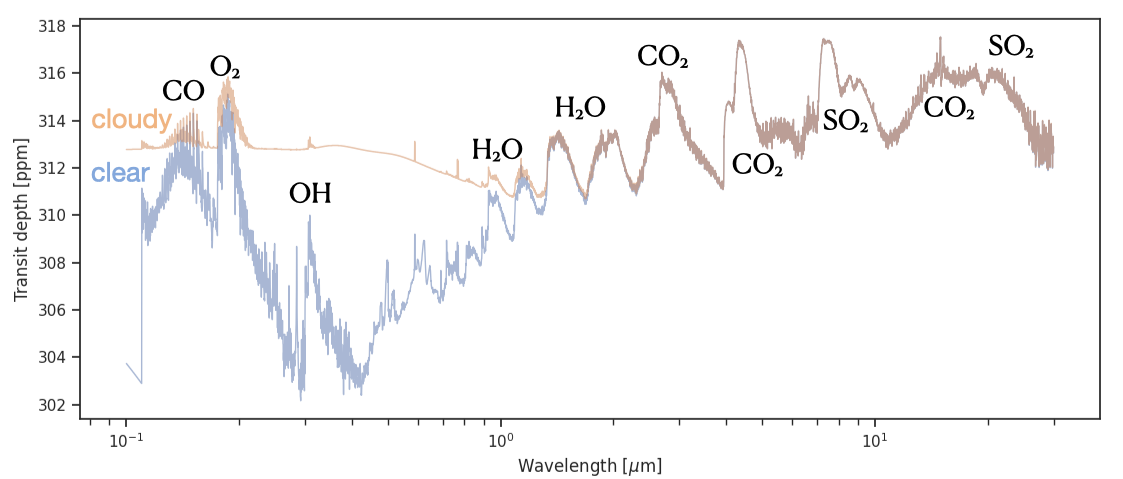}\\
        \caption{Transmission spectrum of the \ce{O2}/\ce{CO2} atmosphere, model 5. Shown are features between $0.1-30$\,$\mu$m. The blue spectrum is the outcome if cloud opacities are omitted. The orange curve shows the full spectrum with clouds. Particularly the UV and optical are affected by cloud opacities. Major gas contributors to the spectrum are \ce{CO2}, \ce{H2O}, \ce{SO2}, \ce{O2}, \ce{CO} and \ce{OH}.}
\label{fig:exolyn_spectrum}
\end{figure*}

Figure \ref{fig:exolyn_spectrum} shows the transmission spectrum resulting from model $5$, the most oxidised of the five cases discussed in section \ref{obudget}. The orange model shows the cloudy spectrum with the mixed grain cloud model and the blue curve is the transmission spectrum without any clouds. Both are generated with petitRADTRANS as described in section \ref{pRT}. First, it is to note, that the atmospheric scale height of this atmosphere is small, with a mean molecular weight of $\mu=34$. The features are almost non existent, on the order of $5$\,ppm in the JWST MIRI wavelength range. Beyond $1$\,$\mu$m, the cloudy and clear spectra are identical. The infrared is completely dominated by \ce{CO2}, \ce{CO}, \ce{H2O} and \ce{SO2} absorption. Since the cloud grain is primarily composed of \ce{TiO2(s)} mixed with \ce{FeO(s)} and \ce{SiO2(s)}, one could expect to see some of the characteristic infrared features from Figure \ref{nk} in the spectrum. However, none of them are visible. The reason for this are first of all the high abundances of \ce{CO2}, \ce{CO}, \ce{H2O} and \ce{SO2} in this particular atmosphere, leading to very strong opacities of these species in the infrared, see the mass mixing ratios in Figure \ref{fig_Crich_oxi5}. On the other hand, the cloud opacities in the infrared are in general weaker compared to the UV and optical. This is shown in Figure \ref{nk}, where for a $0.01$\,$\mu$m large particle the extinction coefficient drops with wavelength. The \ce{Ti}-\ce{O} band has a unique refractive index drop in the UV between $0.2-0.5$\,$\mu $m which can lead to features in the UV and optical slope. However, in this atmosphere, this feature does not show either. The major effect of the clouds in the UV and optical is to dampen the features from gas species such as \ce{CO}, \ce{O2} and \ce{OH} between $0.1-1$\,$\mu$m.

\begin{figure*}
    \centering
    
        \centering
        \includegraphics[width=0.98\textwidth]{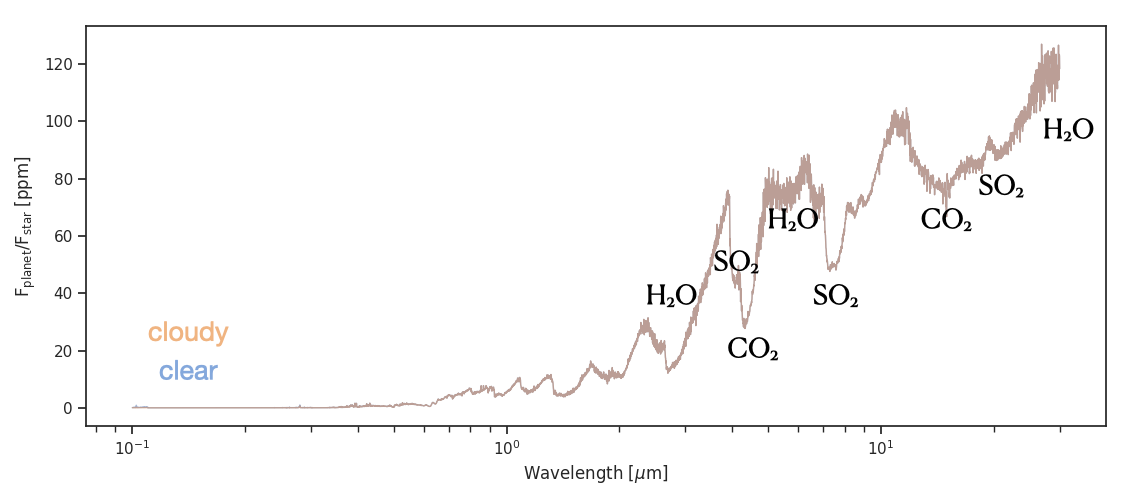} \\
        \caption{Emission spectrum of the \ce{O2}/\ce{CO2} atmosphere, model 5. Shown are features between $0.1-30$\,$\mu$m. The blue spectrum is the outcome if cloud opacities are omitted. The orange curve shows the full spectrum with clouds. Major gas contributors to the spectrum are \ce{CO2}, \ce{H2O} and \ce{SO2}.}
\label{fig:spectrum_emission}
\end{figure*}

The slope of the spectrum in the UV and optical could potentially be an indicator of the cloud species. However, it is strongly affected by the mean molecular weight of the atmosphere, the cloud species and their particle sizes \citep{Pinhas2017,Wakeford2017}. These atmospheric properties need to be disentangled first before concluding on one particular cloud species from those spectral properties. Hence, one would need to look at infrared features to distinguish titanium oxides from each other and from other cloud species. However, as discussed this is not possible for this particular case, due to the infrared features being masked by features of gaseous species as well them being affected by the weakening of extinction properties of clouds with wavelength.

In general, the two spectra show that in such an atmosphere the mean molecular weight is extremely high and features from the gas phase and from the clouds are weak. It is to note that even though the cloud in the \ce{O2}/\ce{CO2} atmosphere extends through six orders of magnitude in pressure, it does not flatten any spectral features in the infrared. In emission, the clouds have no features at all, see Figure \ref{fig:spectrum_emission}. These results are likely impacted by our modelling choice. We are not accounting for the effect of clouds on the temperature structure, which is an essential component for the emission. Furthermore, we are using the same temperature-pressure profile for the emission and the transmission spectrum with a heat redistribution factor $f=\frac{1}{3}$. In reality, we would expect the terminators to be colder and  the sub-stellar point to be hotter, enhancing cloud formation in the regions probed by transmission.

\subsubsection{Spectral imprints of clouds in a \ce{C}- atmosphere}\label{spectra_CH}

\begin{figure*}
        \centering
        \includegraphics[width=0.98\textwidth]{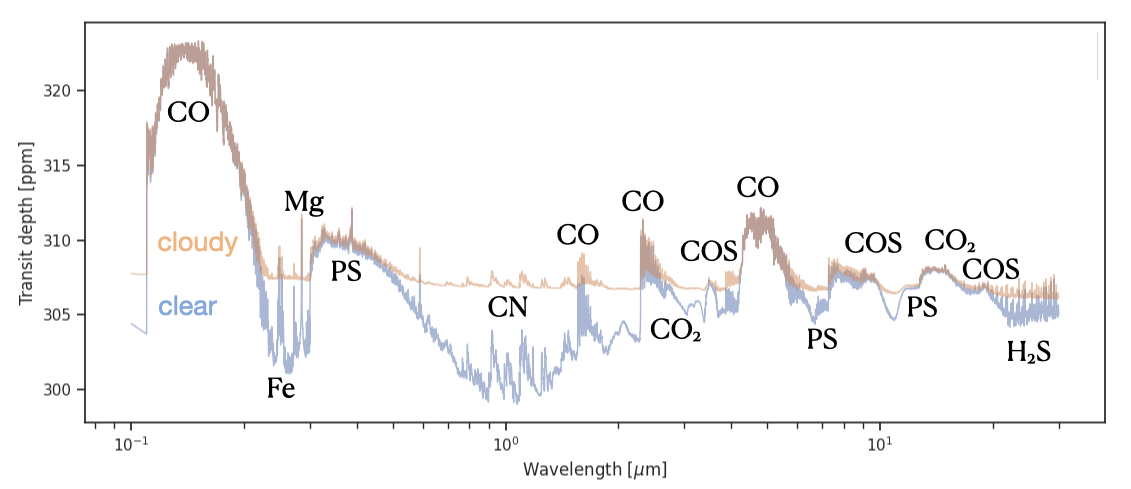}
        \caption{Transmission spectrum of the \ce{C}-atmosphere, model 5 with \ce{C}/\ce{O}=2. Shown are features between $0.1-30$\,$\mu$m. The blue spectrum is the spectrum if cloud opacities are omitted. The orange spectrums shows the full spectrum with clouds. The spectrum is strongly affected by graphite clouds through the entire wavelength range with a flattening effect.}
\label{fig:spectrum_emission_C}
\end{figure*}

In the transmission spectrum of the \ce{C}-atmosphere, spectral features mainly originate from \ce{CO}, \ce{PS} and some \ce{CO2}, \ce{COS} and \ce{H2S}. The graphite cloud does not generate characteristic features but has a strong flattening effect on the gaseous features. This is due to the high and featureless extinction coefficient of graphite, see Figure \ref{nk}. Even though \ce{C}/\ce{O} ratios of the gas phase can be similar for \ce{C}-atmospheres with bulk \ce{C}/\ce{O}$>1$ and \ce{CO}-atmospheres with bulk \ce{C}/\ce{O}$\approx1$, flat spectral features can point towards high bulk \ce{C}/\ce{O} and to the formation of graphite clouds.

Graphite condensing on top of the silicate mantle has been suggested before by \citet{Kuchner2005}. The authors predict that such a carbon rich planet could point towards planet formation having taken place in locally carbon enriched environments of the disk. Hence detecting graphite clouds could provide us with some hints towards understanding disk dynamics and formation theory.

\subsection{The effect of vertical mixing on cloud extent and composition}\label{eddydiff}

Vertical mixing is highly unknown, especially for hot, rocky planets. It is commonly parameterised by the eddy diffusion parameter K$_{\text{zz}}$.  In gas giants, K$_{\text{zz}}$ can span values between K$_{\text{zz}}=10^{3}$\,cm$^2$ s$^{-1}$ to K$_{\text{zz}}=10^{12}$\,cm$^2$s$^{-1}$ \citep{Venot2018}. So far, we have been using a value of K$_{\text{zz}}=10^{10}$\,cm$^2$s$^{-1}$ in all of our models to investigate cloud formation with efficient gas replenishment. In this section, we explore the relevance that this parameter might have in the determination of the clouds by comparing our results with this strong eddy diffusion to the cases with weaker vertical mixing of K$_{\text{zz}}=10^{8}$\,cm$^2$s$^{-1}$ and K$_{\text{zz}}=10^{6}$\,cm$^2$s$^{-1}$.

\begin{figure*}
    \centering
    \includegraphics[width=0.45\textwidth]{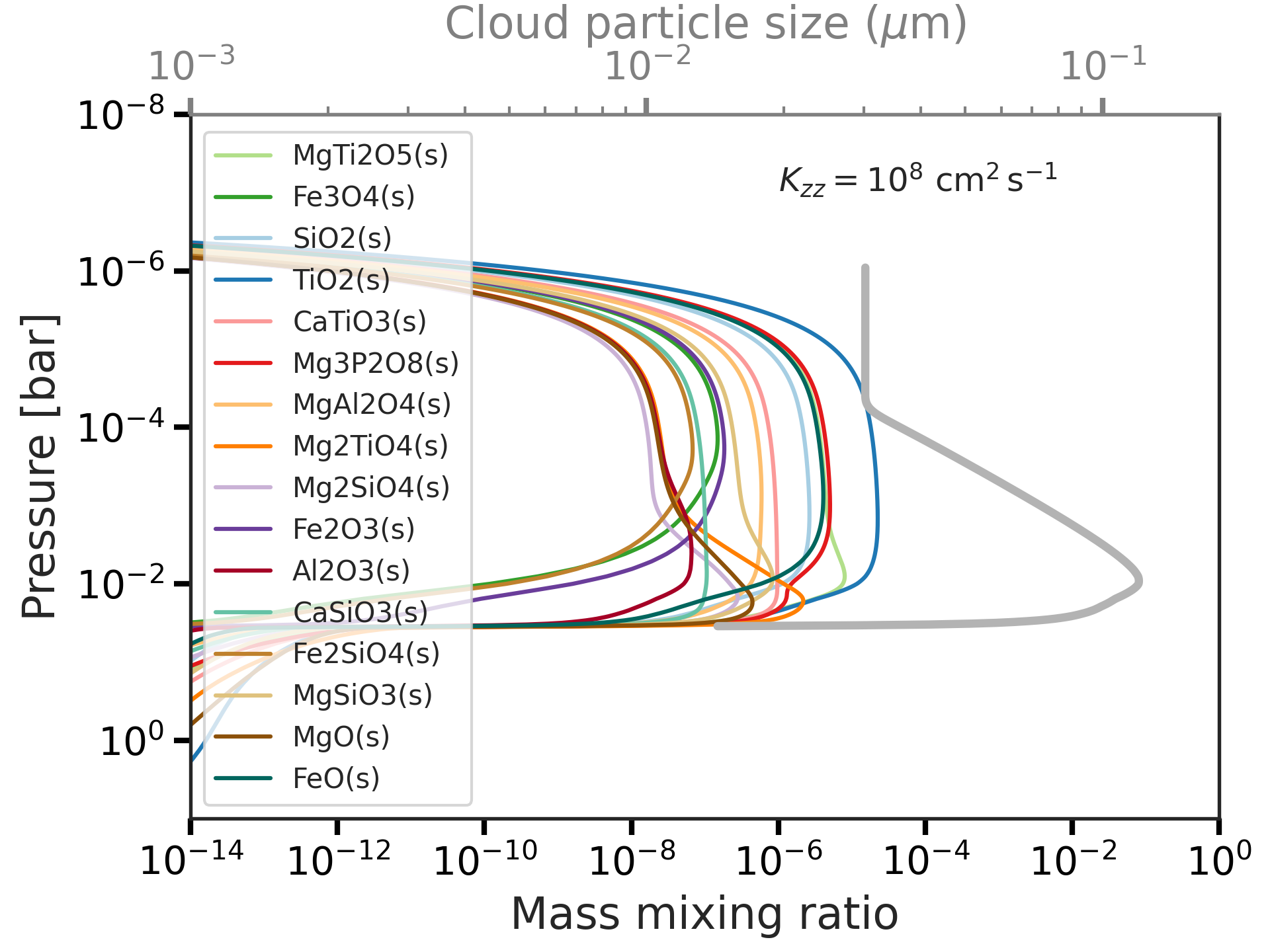}
    \includegraphics[width=0.45\textwidth]{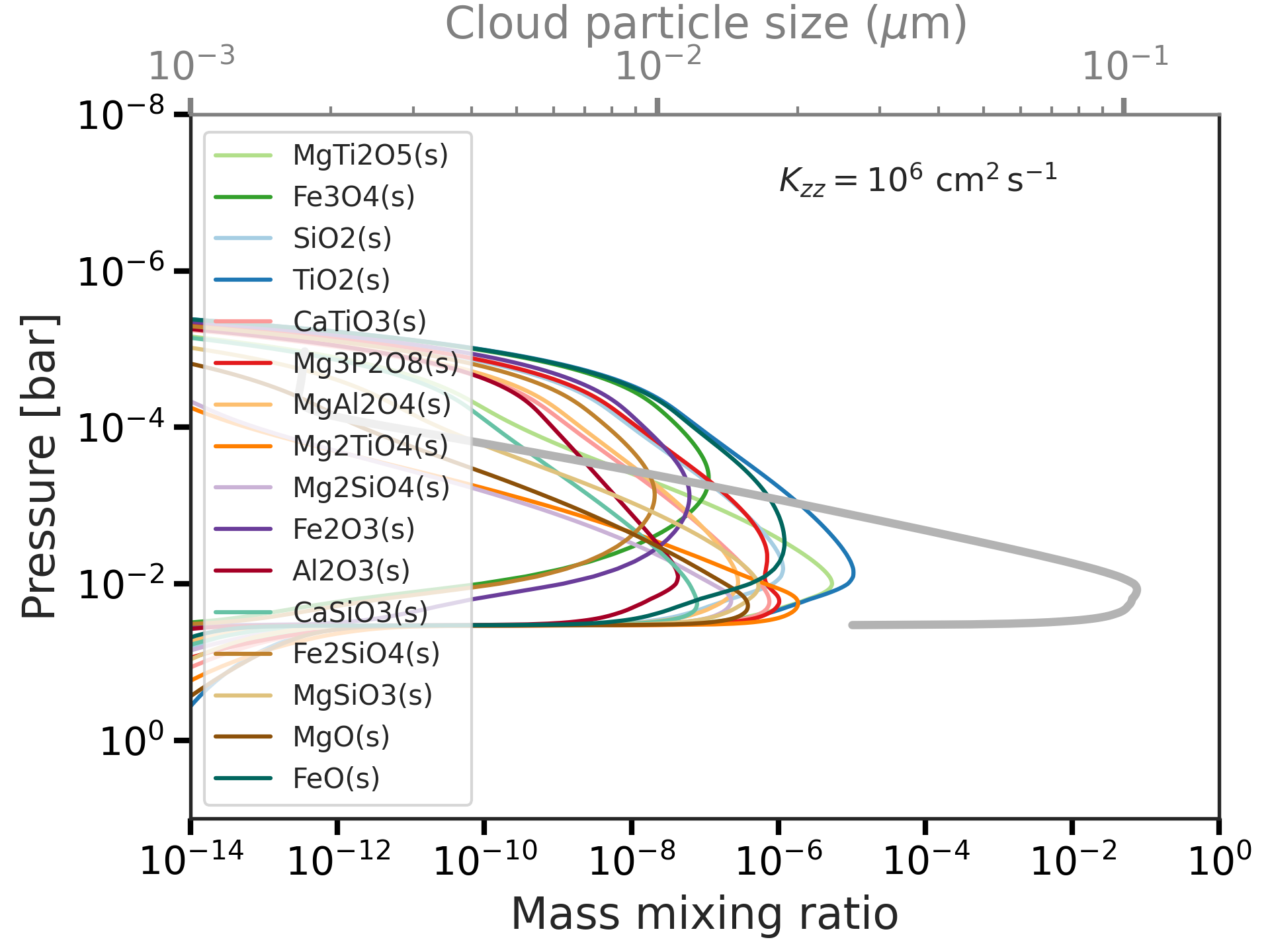}
\caption{Shown are mass mixing ratios of the condensate/cloud mass mixing ratios in the atmosphere of model $5$. We have computed them with {\textsc{ExoLyn}} and {\textsc{FastChem}}3 combined as described in section \ref{exolyn}. The left hand side panel shows the cloud for K$_{\text{zz}}$=$10^8$\,cm$^{2}$s$^{-1}$ and the right hand side panel for K$_{\text{zz}}$=$10^6$\,cm$^{2}$s$^{-1}$.}
\label{Oxi5_Kzz}
\end{figure*}

Figure \ref{Oxi5_Kzz} shows the composition of the cloud in model $5$ for different mixing strengths. As mixing decreases, the cloud becomes smaller in extent and less uniform. While for K$_{\text{zz}}=10^{10}$\,cm$^2$ s$^{-1}$, the cloud extends from below $0.01$\,bar to $5\cdot10^{-8}$\,bar, it evaporates at $10^{-6}$\,bar if K$_{\text{zz}}=10^{8}$\,cm$^2$ s$^{-1}$ and at $10^{-5}$\,bar if K$_{\text{zz}}=10^6$\,cm$^2$ s$^{-1}$. The composition and the particle sizes of the cloud grains also change with mixing strength. At the smallest mixing of K$_{\text{zz}}=10^6$\,cm$^2$ s$^{-1}$, the layer structure of the cloud is particularly distinct. A bottom layer forms particles of $0.1$\,$\mu$m between $0.05$\,bar - $10^{-2}$\,bar. It is mainly composed of \ce{Mg2TiO4(s)}, \ce{TiO2(s)} and \ce{Mg3P2O8(s)}. Above this layer is a \ce{TiO2(s)} cloud layer mixed with some \ce{MgTi2O5(s)} extending from $10^{-2}$\,bar to $5\cdot10^{-3}$\,bar with particles of sizes between $0.05$\,$\mu$m to $0.1$\,$\mu$m. Above these pressures, iron oxides compose large fractions of the cloud. \ce{FeO(s)} condenses first followed by \ce{Fe3O4(s)}. From  $3\cdot10^{-5}$\,bar on, the cloud is composed of primarily \ce{FeO(s)}, \ce{TiO2(s)} and \ce{Fe3O4(s)} with small particles $<0.02$\,$\mu$m. This three layer structure is a similar structure to the one in the rainout model. It forms at weak mixing only because an inefficient transport of material means that solid species stay in the regions where they are thermodynamically stable. The particle size is largest towards the bottom of the cloud, where setteling is the strongest. The cloud becomes thinner with altitude and the particles shrink. Overall, the species which condenses the most efficiently onto the cloud grains is \ce{TiO2(s)}. Setteling of \ce{CaTiO3(s)} and \ce{Mg2TiO4(s)} in the lower cloud layer is also extremely efficient. However, their abundances drop quickly with altitude.
This separation into three layers disappears at K$_{\text{zz}}=10^8$\,cm$^2$ s$^{-1}$. Only the lowest Mg-oxide layer can be distinguished from a more extended upper layer, where \ce{TiO2(s)} is the major component mixed with other components including silicates and iron oxides. The cloud particles reach their maximal grain sizes of $\approx 0.1$\,$\mu$m around $0.01$\,bar. Towards the cloud top, the particles shrink to $0.03\,\mu$m diameter. The overall particle size increases with  K$_{\text{zz}}$, because the gas replenishment becomes faster leading to quicker particle growth, see Figure \ref{exolyn}. In the case of strongest mixing K$_{\text{zz}}=10^{10}$\,cm$^2$, grain sizes vary between $0.23$ and $0.13$\,$\mu$m.

\begin{figure*}
        \centering\includegraphics[width=0.98\textwidth]{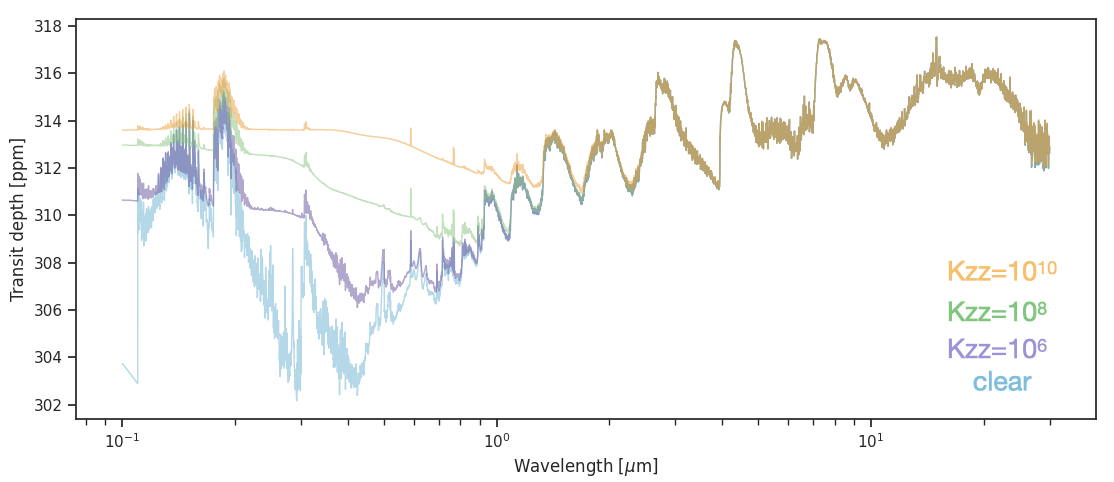}\\
        \caption{Transmission spectrum of the \ce{O2}/\ce{CO2} atmosphere Figure \ref{fig_piecharts} model $5$. Shown are features between $0.1-30$\,$\mu$m. The blue spectrum is simulated without any cloud opacities. The orange, green and purple spectra are cloudy with different strengths of eddy diffusion,  K$_{\text{zz}}=10^{10}$\,cm$^2$ s$^{-1}$, K$_{\text{zz}}=10^8$\,cm$^2$ s$^{-1}$ and K$_{\text{zz}}=10^6$\,cm$^2$ s$^{-1}$ respectively. For each eddy diffusion parameter, the corresponding clouds are shown in sections \ref{O2_comparison} and \ref{eddydiff}. The stronger the vertical mixing, the stronger the cloud affects the spectrum.}
        \label{spectra_Kzz}
\end{figure*}

Figure \ref{spectra_Kzz} illustrates how the differences in the cloud structure due to changes in vertical mixing affect the transmission features of the cloud. The cloud features become weaker with smaller K$_{\text{zz}}$ because the entire cloud is optically thinner. The drop off in the optical slope happens at shorter wavelengths because the small particles in weakly mixed atmospheres scatter shorter wavelengths compared to the larger particles in highly mixed atmospheres.

\section{Discussion}
\subsection{Outgassed atmospheres}
The ability to form a certain condensate depends on the availability of the limiting species. Increasing its abundance can increase the condensate abundance and the condensation temperature. Therefore, it depends on the availability of refractory species which clouds we can expect. \citet{vBuchem2023,vBuchem2024,Zilinskas2022} show that outgassing from a molten surface can enhance the gas phase abundances of \ce{SiO}, \ce{SiO2} and \ce{TiO}. 
This would lead to the formation of atmospheres as predicted by \citet{Zilinskas2022, Piette2023,Seidler2024} where \ce{SiO} strongly affects the temperature structure. \citet{Zilinskas2022} find that the three species are key in forming the temperature structure in outgassed atmospheres and can give rise to strong features in emission spectra. They highlight that \ce{SiO} can be used as a tracer of magma oceans. These molecules can be the limiting species for forming silicon- or titanium condensates and therefore their abundance impacts to what extent cloud formation can take place in such atmospheres. The other way around, cloud formation can deplete \ce{SiO}, \ce{SiO2} and \ce{TiO} in planetary atmospheres and therefore potentially diminish their effect on temperature structure and observability. We will investigate the extent of this effect in a future study.

Within the scope of what we have investigated in this work, we find that in the range of considered compositions, \ce{O}-dominated atmospheres are particularly prone towards cloud formation due to their low temperatures at high altitudes. \citet{Gkouvelis2025} show that due to their higher mean molecular weight, \ce{O2}-atmospheres around small planets are also less likely to escape compared to \ce{H2}-atmospheres. \citet{Cherubim2025} further suggest a trend in sub-Neptunes and super-Earths that smaller planets are more oxidised compared to larger planets. If a sub-Neptune has a \ce{H}-rich envelope, this envelope pushes water to be trapped in the magma ocean \citep[e.g.][]{Kite2019,Kite2020,Dorn&Lichtenberg2021,Gaillard2022,Sossi2023}. After the loss of this envelope, \ce{H2O} is outgassed and the gaseous hydrogen escapes leaving an oxidised atmosphere behind. Therefore, we would expect these small, evolved planets to be particularly relevant for the observations of clouds on strongly irradiated planets.
In a similar way to \ce{H2O}, we also expect that the species which rain out at the surface of the atmosphere would dissolve into the melt to some extent. Our current models assume that solid and liquid species which rain out stay on the surface. Outgassing codes like LavAtmos \citep{vBuchem2023,vBuchem2024} can model gas-magma interactions without considering dissolution into the melt. Integrating the latter for a wide range of rained out condensates would be novel and crucial to consider in the future in order to fully understand the physics and chemistry of cloudy super-Earths and sub-Neptunes.

\subsection{55 Cnc e}
Clouds can also have a cooling effect on the atmosphere. \citet{Loftus2024} propose a cyclical pattern between cooling by clouds and reheating of the atmosphere after cloud dissipation to be a potential source for the observed variability on 55 Cnc e \citep{Demory2016}. Our results suggest that this would mainly happen in the case of an oxygen rich atmosphere, since these are the atmospheres most prone to cloud formation. However, further analysis is needed to confirm this. We will investigate the role of clouds for potential variability in a future work.

We have shown that one likely cloud composition on 55 Cnc e are titanium clouds. However, depending on the surface pressure, on the composition of a potential primary volatile envelope as well as on the magma composition, the available cloud materials can differ drastically from what we find with solar refractory abundances \citep{Gaillard2014,vBuchem2023,vBuchem2024}. Thus, clouds on magma oceans planets like 55 Cnc e could be of a much greater variety. It remains to be discovered, whether 55 Cnc e hosts clouds and if so, what they are made of.

\subsection{Availability of optical data}
So far, research on cloud formation above $1500$\,K has been predominantly focusing on hydrogen-dominated atmospheres [e.g.]\citep{Powell2024,Helling&Woitke2006,Helling2008,Wakeford2015, Helling2017,Huang2024,Bell2024}.
\citet{Helling2008} list reaction rates for \ce{TiO2(s)}, \ce{CaTiO3(s)}, \ce{Al2O3(s)}, \ce{Mg2SiO4(s)}, \ce{MgSiO3(s)}, \ce{MgO(s)}, \ce{Fe2O3(s)}, \ce{FeS(s)}, \ce{FeO(s)}, \ce{Fe(s)}, \ce{SiO(s)}and  \ce{SiO2(s)}. In \citet{Helling2017} additional reaction pathways for carbon bearing species \ce{C(s)}, \ce{SiC(s)}, \ce{TiC(s)} and for \ce{KCl(s)} and \ce{MgS(s)} are available. \citet{Wakeford2017} focus on high temperature condensates classifying them into \ce{Ti}- bearing and \ce{Al}-bearing species. Our findings suggest that further cloud species beyond those mentioned can be relevant. For example, \ce{MgTi2O5(s,l)} and \ce{Mg2TiO4(s,l)} can become major components of cloud particles in \ce{O}-dominated atmospheres or in \ce{H}- and \ce{N}-dominated atmospheres with \ce{C}/\ce{O}$<1$, although, we don't find \ce{Mg2TiO4(s,l)} with abundances $>1$\,ppb in \ce{N}-dominated atmospheres with \ce{C}/\ce{O}$=0.5$. In addition, \ce{Fe3O4(s,l)} needs to be considered for \ce{O}- and \ce{N}-dominated atmospheres with \ce{C}/\ce{O}$<1$.
Recently, also phosphorus has received more attention. \citet{Zilinskas2025} mention the possibility that 55 Cnc e could have a phosphorus rich atmosphere. We find that an atmosphere with abundant sulphur and phosphorus is not a favourable environment for condensation if it is \ce{H}-dominated. However, in \ce{O}-atmospheres which are rich in sulphur and phosphorus, \ce{PO2} can condense to form \ce{(P2O5)2(s)} or \ce{Mg3(PO4)2(s)} if magnesium is available. \ce{Mg3(PO4)2(s)} can also form in \ce{N}-atmospheres with low \ce{C}/\ce{O} ratio and low \ce{H}/\ce{O} ratio. No measured refractive indices for condensate species with \ce{P}-\ce{O} stretch are available.
Thus, more optical data is needed for more accurate predictions of the spectral features of clouds in oxygen rich atmospheres.
In addition, there is also a lack data for high \ce{C}/\ce{O} atmospheres. In some of these \ce{N}-atmospheres, \ce{TiN(s)} and \ce{AlN(s)} clouds can form, but no optical data is available for species with \ce{Ti}-\ce{N} or \ce{Al}-\ce{N} features yet. Furthermore, optical constants used to identify spectral features of cloud particles are a function of temperature \citep{Luna2021}. The data we use has been taken at temperatures $<1000$\,K. However, we consider much hotter atmospheres with temperatures ranging up to $3000$\,K at the surface. Optical data for such hot regimes is not available. This adds uncertainty to the prediction of the features for all the considered materials. Similarly, some solid materials also have different temperature, pressure and composition dependent morphologies which we don't account for, as mentioned in section \ref{kcoefficients}. To interpret spectral features in atmospheres of hot sub-Neptunes and super-Earths correctly it is advantageous to have optical constants ranging from the UV to the infrared at temperatures at least up to 3000\,K for different morphology types of all major cloud materials mentioned in this work.

\subsection{Spectral features}

In section \ref{spectra_oxygen} we show the spectrum for one specific case of \ce{O2}/\ce{CO2} atmospheres. However, the spectral features which will appear are very dependent on the atmospheric composition and the temperature structure.
We expect \ce{O2}-dominated atmospheres to have small scale heights. If refractories are outgassed, this could increase the atmospheric weight even further. In this case, a flattening effect of clouds could be degenerate with a very heavy atmosphere. Features in transmission will be difficult to see. Given the fact that we expect most small planets to have oxygen rich atmospheres \citep[e.g.][]{Cherubim2025}, this makes the characterisation of their atmospheres difficult. We will investigate observability further in future work.

\subsection{Model caveats and future improvements}

In addition to better optical data, our model itself could also benefit from some refinements in the future. So far, our model accounts for gas phase chemistry and condensation reactions but no solid-solid reactions. Common materials we find in the solar environment medium are forsterite and enstatite enhanced with iron \citep{Jaeger1994}. We only consider fayalite, pure forsterite, pure enstatite and iron-oxides \ce{FeO(s)}, \ce{Fe2O3(s)} and \ce{Fe3O4(s)}. Solid-solid reactions could potentially be the cause for the formation of further materials and alter the relative abundances of already present species.

The effect of scattering and absorption by aerosol particles on the temperature structure is another major mechanism which also needs to be considered to understand the feedback between clouds, atmosphere and surface-atmosphere interactions. Investigating this is beyond the scope of this paper but remains a source of uncertainty in our models. 

It is also to note that we have chosen a particular irradiation of the planet which correspond to the irradiation received by 55 Cnc e scaled with a heat redistribution factor of $\frac{1}{3}$. The obtained results will be altered if the planetary parameters or the location on the planet (dayside, terminators, nightside) change. In general, planetary terminators tend to be cooler compared to their day sides. This could enhance cloud formation in these regions compared to our predictions. In addition, horizontal atmospheric movements can lead to asymmetries in planetary atmospheres, which further promotes cloud formation at the planet's terminators e.g. \citep{Lee2023}.

\section{Conclusions}

We have developed a model which couples gas-phase chemistry with radiative transfer and a cloud formation model to investigate the types of clouds that may form in the atmospheres of hot ($>2000$ \,K) sub-Neptunes and super-Earths. We conclude the following:

\begin{enumerate}
      
\item Depending on the composition, different molecules acting as heating agents can form in oxygen poor atmospheres. These are mainly \ce{TiO}, \ce{CN} - if nitrogen and carbon are  abundant -, and \ce{PS} - if phosphorus and sulphur are abundant. These species are strong absorbers in the optical and UV ranges and can cause temperature inversions, inhibiting cloud formation.

\item Oxygen rich atmospheres (\ce{O}-dominated or oxygen enriched \ce{H}- and \ce{N}-atmospheres with \ce{C}/\ce{O}$<1$) are more conducive to cloud formation. The presence of oxygen enables the formation of a wider range of condensates, and these atmospheres are generally cooler, presenting better conditions for cloud formation.

\item A detailed cloud model is essential for accurate cloud characterisation. Equilibrium condensation with rainout underestimates both the cloud mass and the vertical extent of clouds, particularly those forming at lower altitudes. 

\item Vertical mixing has a substantial impact on cloud properties. Stronger vertical mixing leads to larger cloud particles, more extended and uniform cloud decks, and more prominent spectral features.

\item \ce{TiO2} is the dominant cloud species expected in a \ce{O2}/\ce{CO2}-dominated atmosphere with a solar refractory element composition.

\item \ce{O2}/\ce{CO2} atmospheres are heavy, making their spectral features weaker. In the infrared, clouds have little impact on the transmission spectrum. However, in the optical and in the UV they can mute features of some gaseous species like \ce{OH}.

\item Cloud species can be grouped according to their spectral features. Their features are the strongest in the UV and optical, where the slope has unique properties. However, it is crucial to account for degeneracies between particle size, mean molecular weight and cloud species to analyse observed features at these wavelengths. \ce{Ti}-\ce{O} species have a unique feature between $0.2-0.3$\,$\mu$m and \ce{Si}-\ce{O} species between $0.1-0.2$\,$\mu$m. \ce{Fe}-species can produce a shallow, high opacity slope, between $0.1-15$\,$\mu$m. Graphite - \ce{C(s)} - has a relatively strong extinction throughout a broad wavelength range from $0.1-30$\,$\mu$m, leading to a prominent flattening effect of graphite clouds in transmission spectra.

\end{enumerate}

In summary, the oxygen abundance of a planetary atmosphere plays a critical role in its thermal structure and cloud formation. Clouds are predominantly present in oxidised atmospheres of hot sub-Neptune sized planets and can span a variety of compositions. Our findings show that we cannot neglect cloud formation in hot atmospheres as clouds leave imprints on atmospheric climates and planetary spectra. To optimise the analysis of future observations, more detailed models including outgassing from a magma ocean and the impact of clouds on the temperature structure as well as further optical data for solid and liquid species are essential.

\section*{Acknowledgements}

Many thanks to Billy Edwards for fantastic advice on histograms.
L.J.J and Y.M acknowledge support from the European Research Council (ERC) under the European Union’s Horizon 2020 research and innovation programme (grant agreement no. 101088557, N-GINE).

\section*{Data Availability}

All data will be available upon request. We conducted our work with open source codes {\textsc{HELIOS}} \url{https://github.com/HELIOS-framework/HELIOS}, {\textsc{FastChem}}3 \url{https://github.com/NewStrangeWorlds/FastChem}, {\textsc{optool}} \url{https://github.com/cdominik/optool}, and {\textsc{ExoLyn}} \url{https://github.com/helonghuangastro/ExoLyn}.

\bibliography{library}



\appendix

\section{Opacity references} \label{ref_opac}

\begin{table*}
\begin{center}
  \caption{Opacities}
  \label{Tab_opacities} 
  \vspace*{-1mm}  
\resizebox{16.5cm}{!}{
\begin{tabular}{ccccc} 
Species & Source & use & Line list & Line List Reference\\
  \hline
  \ce{Al} & DACE/{\textsc{ExoMol}} & RT/TS & VALD/Kurucz & \citet{Ryab_2015, Kurucz_1992} \\
  \ce{AlH} &  {\textsc{HELIOS}}-K/{\textsc{ExoMol}} &RT/TS &AlHambra & \citet{Yurchenko_2018b}\\
  \ce{AlO} &  {\textsc{HELIOS}}-K/{\textsc{ExoMol}} &RT/TS& ATP & \citet{Patrascu_2015} \\
  \ce{C2H2} & DACE/{\textsc{ExoMol}} &RT/TS &aCeTY & \citet{Chubb_2020}\\
  \ce{C2H4} & DACE/{\textsc{ExoMol}} &RT/TS& MaYTY & \citet{Mant_2018}\\  
  \ce{Ca} & DACE/{\textsc{ExoMol}} & RT/TS&VALD/Kurucz & \citet{Ryab_2015,Kurucz_1992}\\
  \ce{CaH} &  {\textsc{HELIOS}}-K/{\textsc{ExoMol}}&RT/TS & MoLLIST & \citet{Li_2012,Bernath_2020}\\
  \ce{CaO} &  {\textsc{HELIOS}}-K& RT& VBATHY & \citet{Yurchenko_2016}\\
  \ce{CaOH} & DACE{\textsc{ExoMol}}& RT/TS & OYT6 & \citet{Owens_2022}\\
  \ce{C2} & DACE{\textsc{ExoMol}} &RT/TS & 8states & \citet{Yurchenko_2018d} \\
  \ce{CH} & DACE &RT& MoLLIST & \citet{Masseron_2014,Bernath_2020}\\
  \ce{CH3} & DACE&RT & AYYJ & \citet{Adam_2019}\\
  \ce{CH4} & DACE/{\textsc{ExoMol}}& RT/TS& YT34to10/MM & \citet{Yurchenko_2017}\\
  \ce{CN} &  {\textsc{HELIOS}}-K/{\textsc{ExoMol}} &RT/TS& Trihybrid/MoLLIST & \citet{Syme_2021}\\ 
  \ce{CO} & DACE/{\textsc{ExoMol}}& RT/TS& Li2015/HITEMP& \citet{Gang_2015}\\
  \ce{CO2} & DACE/{\textsc{ExoMol}} &RT/TS& HITEMP/UCL-4000m& \citet{Rothman_2010,Yurchenko_2020}\\
  \ce{CS} & DACE/{\textsc{ExoMol}} &RT/TS& JnK & \citet{Paulose_2015}\\ 
  \ce{Fe} & DACE/{\textsc{ExoMol}} &RT/TS& VALD/Kurucz & \citet{Ryab_2015,Kurucz_1992}\\
  \ce{FeH} & DACE/{\textsc{ExoMol}} &RT/TS& MoLLIST & \citet{Dulick_2003}\\
  \ce{H2+} & DACE &RT& ADJSAAM & \citet{Amaral_2019}\\
  \ce{H2O} & DACE/{\textsc{ExoMol}}&RT/TS& POKAZATEL & \citet{Polyansky_2018}\\
  \ce{H2O2} & DACE &RT& APTY & \citet{Refaie_2016}\\ 
  \ce{H2CO} & DACE &RTS& AYTY & \citet{Refaie_2015}\\
  \ce{H2S} & DACE/\texttt{ExoMol} &RT& AYT2 & \citet{Azzam_2016}\\  
  \ce{H3O+} & DACE &RT& eXeL & \citet{Yurchenko_2020}\\  
  \ce{HCN} &  {\textsc{HELIOS}}-K/{\textsc{ExoMol}} &RT/TS& Harris & \citet{Barber_2014}\\
  \ce{HNO3} & DACE &RT/TS& AIJS & \citet{Pavlyuchko_2015}\\ 
  \ce{HS} &  {\textsc{HELIOS}}-K/{\textsc{ExoMol}} &RT/TS& GYT & \citet{Gorman_2019}\\  
  \ce{K} & DACE/{\textsc{ExoMol}} &RT/TS& VALD/Kurucz & \citet{Ryab_2015,Kurucz_1992}\\
  \ce{KOH} & DACE &RT/TS& OYT4 & \citet{Owens_2021}\\
  \ce{Mg} & DACE/{\textsc{ExoMol}} &RT/TS& Kurucz & \citet{Kurucz_1992}\\
  \ce{MgH} &  {\textsc{HELIOS}}-K/{\textsc{ExoMol}} &RT/TS& MoLLIST & \citet{GharibNezhad_2013,Bernath_2020}\\  
  \ce{MgO} &  {\textsc{HELIOS}}-K/{\textsc{ExoMol}}&RT/TS& LiTY & \citet{Li_2019}\\
  \ce{N2} & DACE/{\textsc{ExoMol}}&RT/TS& WCCRMT & \citet{Western_2018}\\
  \ce{N2O} & DACE& RT/TS& HITEMP2019/TYM & \citet{Hargreaves_2019}\\
  \ce{Na} & DACE/{\textsc{ExoMol}} &RT/TS& VALD/Kurucz & \citet{Ryab_2015,Kurucz_1992}\\
  \ce{NaH} &  {\textsc{HELIOS}}-K/{\textsc{ExoMol}}& RT/TS& Rivlin & \citet{Rivlin_2015}\\
  \ce{NaO} & DACE &RT& NaOUCMe & \citet{Mitev_2022}\\
  \ce{NaOH} & DACE/{\textsc{ExoMol}}&RT/TS & OYT5 & \citet{Owens_2021}\\
  \ce{NH} & DACE& RT& MoLLIST & \citet{Fernando_2018}\\
  \ce{NH3} & DACE/{\textsc{ExoMol}}&RT/TS& CoYuTe & \citet{Coles_2019}\\ 
  \ce{NO} & DACE&RT & XABC & \citet{Wong_2017,Qu_2021}\\ 
  \ce{NS} & DACE &RT& SNaSH & \citet{Yurchenko_2018c}\\ 
  \ce{OH} & DACE/{\textsc{ExoMol}}&RT/TS& HITEMP/MoLLIST & \citet{Rothman_2010}\\  
  \ce{OH+} & DACE&RT & MoLLIST & \citet{Hodges_2017} \\    
  \ce{PC} & DACE/{\textsc{ExoMol}}&RT/TS & MoLLIST & \citet{Ram_2014,Qin_2021}\\    
  \ce{PH} &  {\textsc{HELIOS}}-K/{\textsc{ExoMol}} &RT/TS& LaTY & \citet{Langleben_2019}\\ 
  \ce{PH3} & DACE/{\textsc{ExoMol}} &RT/TS& SAlTY & \citet{Sousa_2015}\\ 
  \ce{PO} & DACE/{\textsc{ExoMol}}&RT/TS & POPS & \citet{Prajapat_2017}\\ 
  \ce{PS} &  {\textsc{HELIOS}}-K/{\textsc{ExoMol}} &RT/TS& POPS & \citet{Prajapat_2017}\\ 
  \ce{S} & DACE &RT& VALD & \citet{Ryab_2015}\\ 
  \hline
\end{tabular}}
\end{center}
\vspace*{-2mm}
{Each column gives source linelist, use and the reference for the opacities used in the radiative transfer(RT)/ transmission spectrum (TS). The RT opacities are as in \citet{Zilinskas2023}. The TS opacities are {\textsc{petitRADTRANS}} opacities.}
\end{table*}

\begin{table*}
\begin{center}
  \caption{Opacities}
  \label{Tab_opacities_2} 
  \vspace*{-1mm}
\resizebox{16.5cm}{!}{
\begin{tabular}{ccccc} 
Species & Source & use & Line list & Line List Reference \\
  \hline
  \ce{Si} & DACE/{\textsc{ExoMol}} & RT/TS & VALD/Kurucz & \citet{Ryab_2015,Kurucz_1992}\\
  \ce{SiH} &  {\textsc{HELIOS}}-K/{\textsc{ExoMol}} & RT/TS & Sightly & \citet{Yurchenko_2018a}\\
  \ce{SiH2} &  {\textsc{HELIOS}}-K/{\textsc{ExoMol}}& RT/TS & CATS & \citet{Clark_2020}\\ 
  \ce{SiH4} & DACE/{\textsc{ExoMol}} & RT/TS & OY2T & \citet{Owens_2017}\\  
  \ce{SiO} &  {\textsc{HELIOS}}-K/{\textsc{ExoMol}}& RT/TS & SiOUVenIR & \citet{Yurchenko_2022}\\
  \ce{SiO2} & DACE/{\textsc{ExoMol}}& RT/TS & OYT3 & \citet{Owens_2020}\\
  \ce{SiS} & DACE/{\textsc{ExoMol}}& RT/TS& UCTY & \citet{Upadhyay_2018}\\
  \ce{SiN} & DACE & RT/TS & SiNfull & \citet{Semenov_2022}\\   
  \ce{SO} & DACE/{\textsc{ExoMol}} & RT/TS & SOLIS & \citet{Brady_2024}\\
  \ce{SO2} & {\textsc{ExoMol}} & RT/TS & ExoAmes & \citet{Underwood_2016a}\\ 
  \ce{SO3} & {\textsc{ExoMol}} & RT & UYT2 & \citet{Underwood_2016b}\\ 
  \ce{Ti} & DACE/{\textsc{ExoMol}} & RT/TS & VALD/Kurucz & \citet{Ryab_2015, Kurucz_1992}\\
  \ce{TiH} &  {\textsc{HELIOS}}-K/{\textsc{ExoMol}} & RT/TS & MoLLIST & \citet{Burrows_2005, Bernath_2020}\\
  \ce{TiO} &  {\textsc{HELIOS}}-K/{\textsc{ExoMol}} & RT/TS & Toto & \citet{McKemmish_2019}\\
  \ce{COS} & {\textsc{ExoMol}} & TS & OYT8 & \citet{Owens2024}\\
  \ce{PN} & {\textsc{ExoMol}} & RT/TS & PaiN & \citet{Semenov2025}\\
  \ce{O2} & DACE/{\textsc{ExoMol}} & RT/TS & HITRAN & \citet{Chubb_2020}\\
 \ce{O} & DACE/{\textsc{ExoMol}} & RT/TS & Kurucz & \citet{Kurucz_1992}\\
  \hline 
  \ce{O2}, \ce{CO2}, \ce{CO}, \ce{N2} & & RT/TS & Scattering &  \\ 
  \ce{H2}, \ce{H2O}, \ce{H}, \ce{CH4} & &RT/TS & Scattering & \\
   \hline 
  \ce{H-}  &  & RT/TS &  Continuum (bf $\&$ ff) & \citet{John_1988}, \citet{Gray_2008} \\ 
   \hline 
  \ce{H2-H2} & {\textsc{petitRADTRANS}} & RT/TS & CIA & \citet{Borysow2001, Borysow2002} \\
  \ce{O2-O2} & {\textsc{petitRADTRANS}} & TS & CIA & \citet{Karman2019} \\ 
  \ce{N2-H2} & {\textsc{petitRADTRANS}}& TS & CIA & \citet{Karman2019} \\   
  \ce{N2-He} & {\textsc{petitRADTRANS}} & TS & CIA & \citet{Karman2019} \\
  \ce{H2O-N2} & {\textsc{petitRADTRANS}} & TS & CIA & \citet{Kofman_Vilanueva2021} \\
  \ce{H2O-H2O} & {\textsc{petitRADTRANS}} & TS & CIA & \citet{Kofman_Vilanueva2021} \\
  \hline
\end{tabular}}
\end{center}
\vspace*{-2mm}
{Each column gives source linelist, use and the reference for the opacities used in the radiative transfer(RT)/ transmission spectrum (TS). The RT opacities are as in \citet{Zilinskas2023}. The TS opacities are {\textsc{petitRADTRANS}} opacities.}
\end{table*}


\bsp	
\label{lastpage}
\end{document}